\documentclass[aps,pra,showpacs,twoside,twocolumn,longbibliography,10pt]{revtex4-2}
\usepackage[colorlinks=true, citecolor=blue, urlcolor=blue, linkcolor = blue ]{hyperref}
\usepackage{epsfig,newlfont,amssymb,amsfonts,amsmath,bm,subfigure,palatino,mathtools,amsthm,braket,times,soul,enumitem,color}
\usepackage[normalem]{ulem}
\newcommand{\stkout}[1]{\ifmmode\text{\sout{\ensuremath{#1}}}\else\sout{#1}\fi}
\usepackage[english]{babel}
\usepackage[utf8]{inputenc}
\usepackage{array}
\usepackage{xcolor}
\usepackage{graphics}

\def\Tr{\text{Tr}}

\usepackage{amsthm}
\usepackage{verbatim}
\usepackage{bbm}
\usepackage{wrapfig}

\usepackage{hyphenat}

\newlength\figureheight 
\newlength\figurewidth 

\begin{document}

\title{Comparing physical quantities with finite-precision: beyond standard metrology and an illustration for cooling in quantum processes}

\author{Anindita Sarkar}
\email{anindita2712@gmail.com}

\author{Paranjoy Chaki} 
\email{pcapdj16@gmail.com}

\author{Priya Ghosh} 
\email{priyaghosh1155@gmail.com}

\author{Ujjwal Sen}
\email{ujjwal@hri.res.in}

\affiliation{Harish-Chandra Research Institute,  A CI of Homi Bhabha National Institute, Chhatnag Road, Jhunsi, Prayagraj  211 019, India}

\begin{abstract}
We propose a general framework to compare the values of a 
physical quantity pertaining to two - or more - physical setups, in the finite-precision scenario. 
Such a situation requires us to compare between two ``patches'' on the real line instead of two numbers. Identification of extent of the patches is typically done via standard deviation, as obtained within usual quantum metrological considerations, but can not be always applied, especially for asymmetric error distributions. The extent can however be universally determined by utilizing the concept of percentiles of the probability distribution of the corresponding estimator.  As an application, we introduce the concept of finite-precision cooling in a generic quantum system. 
We use this approach in the working of 
a three-qubit quantum refrigerator governed by Markovian dynamics, and demonstrate the occurrence of cooling within finite precision for both transient and steady-state regimes, across strong- and weak-coupling limits of the inter-qubit interaction.
\end{abstract}
\maketitle
\section{Introduction}

In many physical contexts, one often faces the task of comparing the values of a physical quantity pertaining to  two - or more - different physical setups, such as the temperature of an object before and after refrigeration, the magnetization before and after a phase transition, or the energy of a system before and after extraction, etc. Typically, these quantities are not known in advance and must be estimated. Quantum metrology~\cite{Helstrom1969,HOLEVO1973337,PhysRevLett.72.3439,doi:10.1126/science.1104149,PhysRevLett.94.020502,PhysRevLett.96.010401,Giovannetti2011,holevo,RevModPhys.89.035002,RevModPhys.90.035005,RevModPhys.90.035006,Pirandola2018,PhysRevA.109.L030601,PhysRevA.109.052626,PhysRevA.110.012620,singh2024dimensionalgainsensinghigherdimensional,mondal2024multicriticalquantumsensorsdriven,sahoo2024enhancedsensingstarkweak,PhysRevA.111.042628,Agrawal2025indefinitetime,bhattacharyya2025precisionestimatingindependentlocal,agarwal2025criticalquantummetrologyusing,bhattacharyya2025quantumsensingevenversus,saha2025entanglementconstrainedquantummetrologyrapid,pal2025rolephaseoptimalprobe,agarwal2025quantumsensingultracoldsimulators,mondal2025optimalquantumprecisionnoise,chaki2025nonpositivemeasurementsarentbeneficial} offers a convenient framework for such estimation processes, as it harnesses quantum mechanical principles and resources to achieve the highest attainable precision in parameter estimation. Numerous studies~\cite{PhysRevLett.96.010401,Giovannetti2011,nagata,Gross_2012,Israel2014} have shown that quantum metrology can surpass the ultimate precision bounds achievable by classical strategies. In this paper, our aim is to develop a framework that addresses this general problem, namely the comparison or distinction between two physical quantities within the constraints of finite precision.

Every estimation process is inherently limited by finite precision, causing the estimated values to spread over ``patches” on the real line.
To compare the values of a physical quantity corresponding to two different physical situations within the quantum metrological framework, one must therefore compare two such patches rather than two numbers.
In conventional quantum parameter estimation theory, the standard deviation of an estimator is commonly used as a measure of estimation error. This measure is appropriate when the estimator’s probability distribution is symmetric. However, in many realistic metrological scenarios, the distribution may be asymmetric, in which case the standard deviation does not accurately represent the uncertainty of the estimator.
To overcome this limitation, we propose a universal approach that quantifies the estimation error using the percentiles of the estimator’s probability distribution, defined as the extent of the patches. This method remains valid irrespective of the symmetry of the distribution, thereby providing a robust, distribution-independent, and unified framework for quantifying estimation error across diverse quantum parameter estimation tasks.

As an illustration of our framework, we introduce the concept of finite-precision cooling, which incorporates the notion of cooling within the limitations of finite precision. Cooling in quantum systems or quantum refrigeration~\cite{PhysRevLett.105.130401, Skrzypczyk_2011, PhysRevLett.108.070604, Correa2014, PhysRevE.89.032115, PhysRevE.91.050102, Mitchison_2015, PhysRevE.92.062101, Nimmrichter2017quantumclassical, Mu_2017, PhysRevE.97.062116, Das_2019, Mitchison03042019, PhysRevE.101.012109, PhysRevA.104.042208, PhysRevA.105.022214, PhysRevA.106.022616, PhysRevA.107.032602, ray2023kerrtypenonlinearbathsenhance, ghosh2024measurementbasedquditquantumrefrigerator, PhysRevA.111.012209, mondkar2025quantumrefrigeratorembeddedspinstar, mondal2025mpembaeffectselfcontainedquantum} provides a natural setting for such an analysis, as it involves comparing two temperatures to determine whether a system has cooled.
Specifically, we apply our finite-precision cooling framework to the three-qubit quantum refrigerator model governed by Markovian dynamics~\cite{Mitchison_2015}. The qubit temperatures are estimated using the minimum-variance unbiased (MVU) estimator, and finite-precision effects are quantified through the percentiles of the estimator’s probability distribution. This analysis is carried out in both the strong- and weak-coupling regimes of the inter-qubit interaction, revealing cooling behavior in both transient and steady-state limits under finite precision. The probability distribution function of the estimator is obtained using the principle of maximum entropy~\cite{PhysRev.106.620, TAGLIANI1998157, 10.1063/1.4819977}.
Thus, our results establish a universal framework for analyzing finite-precision effects in the comparison of physical quantities and provide a statistical foundation for understanding cooling dynamics in quantum systems.

 The rest of the paper is organized as follows. Sec.~\ref{prelim} is devoted to the preliminary concepts which serve as the essential background for our work. Subsec.~\ref{qmet} outlines the basic formalism of temperature estimation in quantum thermometry, whereas subsec.~\ref{qref} describes the three-qubit self-contained quantum absorption refrigerator model, along with the description of the model chosen for the refrigerator dynamics.
 In Section~\ref{met}, we develop a general finite-precision framework for comparing physical quantities under different conditions. Unlike conventional methods based on standard deviation, our approach quantifies estimation error using percentiles of the estimator’s probability distribution. In Sec.~\ref{res}, 
we apply our general framework to quantum refrigeration, where temperatures of the cold qubit are estimated and compared to determine finite-precision cooling.
We numerically demonstrate the occurrence of finite-precision cooling in the three-qubit self-contained absorption refrigerator model governed by Markovian dynamics, focusing on both the strong- and weak-coupling limits of the inter-qubit interaction. In each case, we present examples illustrating transient as well as steady-state cooling within the finite-precision framework. The cooling behavior in the strong-coupling regime is discussed in subsec.~\ref{sc}, while that in the weak-coupling regime is analyzed in subsec.~\ref{wc}. Finally, we conclude in Sec.~\ref{con}.

\section{Setting the stage}\label{prelim}
This section presents the essential background for our work. In subsection~\ref{qmet}, we discuss the basics of quantum thermometry. This is followed by a description of a three-qubit quantum absorption refrigerator in subsection~\ref{qref}.  

\subsection{Quantum thermometry}
\label{qmet}
Quantum thermometry~\cite{Correa2015-thermometry,Paris2016-thermometry,Hofer2017-thermometry,DePasquale2018-thermometry,Mehboudi2019-thermometry,Potts2019-thermometry,Jorgensen2020-thermometry,Hovhannisyan2021-thermometry,Rubio2021-thermometry} is a subfield of quantum metrology that aims to estimate the temperature of a physical system as precisely as possible.
Here, we focus on the frequentist approach rather than the Bayesian approach for quantum thermometry. 
Suppose we wish to estimate the temperature $T$ of a $d$-dimensional quantum system in a state $\rho_T$ corresponding to the Hilbert space $\mathcal{H}_d$. As temperature is not an observable in quantum mechanics, it is estimated from the measurement data obtained by measuring any quantum observable $\Theta$ with temperature-dependent statistics, (say, the energy of the system)~\cite{PhysRevE.83.011109}. This is done by making a suitable measurement $\{\Pi_\theta\}$ on the state $\rho_T$, resulting in outcomes $\{\theta\}$. The measurement operators satisfy $\Pi_\theta\geq 0$ and $\sum_{\theta} \Pi_\theta=\mathbb{I}_d$, where $\mathbb{I}_d$ is the identity operator of the state space corresponding to $\mathcal{H}_d$.  Each outcome $\theta$ is associated with a  probability $p_{T}(\theta)$, which is given by $p_{T}(\theta)\coloneqq\Tr(\Pi_{\theta}\rho_T)$ according to the Born rule~\cite{Born1926} ($\Tr (\cdot)$ denotes the trace of the operator ``$\cdot$"). The estimate is formally obtained from the measurement data using a function $\hat{T}(\theta)$ known as the estimator. As the measurement data are random, the estimator, being a function of data, is itself a random variable. We restrict ourselves to only those estimators that
 satisfy the following conditions for local unbiasedness at $T=T_0$: 
\begin{align}
   \int d\theta \hat{T}(\theta) p_{T_0}(\theta)=T_0, \quad \int d\theta \hat{T}(\theta) \frac{dp_{T_0}(\theta)}{dT}\bigg \vert_{T_0}=1, \label{lub}
\end{align}
where $T_0$ is the true value of the system temperature. Such estimators are called locally unbiased estimators. 
For a given parameter, there can be many possible (locally) unbiased estimators.

 Obtaining the estimate with the best possible precision entails minimizing the error in estimation. In quantum metrology, usually the mean squared error of the estimator is considered, which equals the variance in case of unbiased estimators. The central result in this regard is given by the (Classical) Cram\'er-Rao bound (CCRB)~\cite{MR15748,a61aa5fe-d74a-3133-bed2-f35c3c555015}, which places a lower bound on the variance of any unbiased estimator by minimizing the variance over all possible (unbiased) estimators for a fixed (but arbitrary) choice of state and measurement:
\begin{align}
    \Delta^2 \hat{T}\geq\frac{1}{NF_C}.\label{ccrb}
\end{align}
Here, $\Delta^2 \hat{T}$ is the variance of the estimator $\hat{T}(\theta)$ and $N$ denotes the number of repetitions of the experiment performed on the system. The quantity $F_C$ is known as the (classical) Fisher information and it is defined as
\begin{equation*}
    F_C\coloneqq\int d\theta p_T(\theta)\left(\frac{d \ln p_T(\theta)}{dT}\right)^2.
\end{equation*}
The saturability of the CCRB is given by the condition~\cite{10.5555/151045}
\begin{equation}
    \frac{d\ln p_T(\theta)}{dT}=F_C\left(\hat{T}(\theta)-T\right). \label{satur}
\end{equation}
The unbiased estimators whose variances satisfy the CCRB are referred to as minimum-variance unbiased (MVU) estimators. The CCRB can be tightened further by minimizing the variance over all possible measurements, resulting in the quantum Cram\'er-Rao bound (QCRB)~\cite{PhysRevD.23.357,PhysRevD.23.357,PhysRevLett.72.3439,holevo}
\begin{align}
    \Delta^2 \hat{T}\geq\frac{1}{NF_C}\geq\frac{1}{NF_Q},\label{qcrb}
\end{align}
where $F_Q$ is the quantum Fisher information (QFI) satisfying $F_Q \geq F_C$. $F_Q$ can always be attained from the measurement statistics where the measurement is performed in the eigenbasis of a special operator known as the symmetric logarithmic derivative (SLD) operator $\Lambda_T$. The SLD operator $\Lambda_T$ is implicitly defined via the equation
\begin{equation}
    \frac{d\rho_T}{dT}=\frac{1}{2}\left(\Lambda_T\rho_T+\rho_T\Lambda_T\right).
\end{equation}
Let us denote the eigenvalues and eigenvectors of $\rho_T$ as $\{\lambda_i\}$ and $\{\ket{e_i}\}$ respectively. 
The matrix elements of the SLD operator with respect to $\{\ket{e_i}\}$ is given by
\begin{equation}
    \bra{e_i}\Lambda_T\ket{e_j}=
    \begin{cases}   \frac{2\bra{e_i}\frac{d\rho_T}{dT}\ket{e_j}}{\lambda_i+\lambda_j} &\quad (\lambda_i+\lambda_j\neq0)\\ 0 &\qquad \text{otherwise} .
    \end{cases} \label{sld}
\end{equation}
    The QFI of the state $\rho_T$ can be expressed in terms of the SLD operator $\Lambda_T$ as 
\begin{equation*}
    F_Q\coloneqq\Tr(\rho_T\Lambda^2_T).
\end{equation*}

\subsection{Three-qubit Quantum Refrigerators}
\label{qref}
A quantum refrigerator is a thermal device that is used to cool quantum systems. A typical quantum absorption refrigerator~\cite{PhysRevLett.105.130401,Mitchison_2015} consists of three interacting qubits. Among these three qubits, the first—referred to as the cold qubit, is the object to be cooled, aided by the second and third qubits, known respectively as the work qubit and the hot qubit. The total Hamiltonian of the three-qubit system is given by $\bar{H}_S=\bar{H}_\text{loc}+\bar{H}_\text{int}$, where $\bar{H}_\text{loc}$ is composed of local Hamiltonians of each qubit, while $\bar{H}_\text{int}$ refer to the interaction between the qubits. 
The cold, work, and hot qubits are each independently coupled to separate heat baths at temperatures \( \bar{T}_1 \), \( \bar{T}_2 \), and \( \bar{T}_3 \), respectively, satisfying \( \bar{T}_1 \leq \bar{T}_2 < \bar{T}_3 \).
 
The refrigerator is self-contained~\cite{PhysRevLett.105.130401, Skrzypczyk_2011} if it can work without external energy sources. The qubits are assumed to be initially in thermal equilibrium with their respective baths. Thus, the initial state of the $j$th qubit is given by $\bar{\tau}_j \coloneqq \frac{\exp(-\bar{\beta}_j \bar{H}_j)}{\Tr[\exp(-\bar{\beta}_j \bar{H}_j)]}$ with $\bar{\beta}_j\coloneqq\frac{1}{\bar{T}_j}$ and $\bar{H}_j$ denotes the local Hamiltonian of the $j$th qubit ($j=1,2,3$). 
Consequently, the initial state of the refrigerator is given by $\bar{\rho}_0 \coloneqq \bar{\tau}_1 \otimes \bar{\tau}_2 \otimes \bar{\tau}_3$.

At time $\bar{t}=0$, the interaction between the qubits is turned on. The state of the three-qubit system $\bar{\rho}(\bar{t})$ at any instant $\bar{t}$, is obtained by solving the master equation
\begin{align}
    \frac{\partial\bar{\rho}(\bar{t})}{\partial \bar{t}}=-i[\bar{H}_S,\bar{\rho}(\bar{t})]+\Phi(\bar{\rho}(\bar{t})).
\end{align}
Here, the term $\Phi(\bar{\rho}(\bar{t}))$ is included to account for the effects of heat baths on the refrigerator dynamics. Its exact form depends on the model chosen for thermalization. The refrigerator is able to cool qubit $1$ if the final temperature of the evolved cold qubit, defined by $\bar{T}^f_1(\bar{t})$ satisfy $\bar{T}^f_1(\bar{t}) < \bar{T}_1$ at least some particular instant of evolving time $\bar{t}$. We refer the cooling as transient cooling, reflecting the fact that cooling occurs at transient regime ($\partial\bar{\rho}(\bar{t})/\partial \bar{t}\neq 0$). In contrast, steady-state cooling occurs if the steady-state temperature ($\bar{T}^s_1$) of the cold qubit follows $\bar{T}^s_1<\bar{T}_1$, where the steady state $\bar{\rho}_s$ is the state of the three-qubit refrigerator system for which $\partial\bar{\rho}_s(\bar{t})/\partial \bar{t}= 0$ is satisfied.

In our work, we have taken $\bar{H}_{\text{loc}}=KH_{\text{loc}}$ and $\bar{H}_{\text{int}}=KH_{\text{int}}$,  $K$ being a constant with the units of energy. The dimensionless Hamiltonians $H_{\text{loc}}$ and $H_{\text{int}}$ are taken in the following way: 
\begin{equation}
    H_{\text{loc}}=\frac{1}{2}\sum_{j=1}^3 E_j\sigma_j^z,\; H_{\text{int}}=g(\ket{010}\bra{101}+\ket{101}\bra{010}).
\end{equation}
Where $E_j>0\;\forall\;j$. The operator $\sigma_j^z$ is the Pauli-$z$ matrix, acting on the $j$th qubit.  The states $\ket{0}$ and $\ket{1}$ respectively denote the ground and excited states of each qubit.  The parameters $\{E_j\}$ and $g$ refer to the dimensionless energy gap between the energy levels of each qubit and the interaction strength between the qubits, respectively. We also introduce  dimensionless temperatures $T_1,T_2,T_3$, related to the initial temperatures of the three qubits as $T_j\coloneqq\frac{K_B}{K}\bar{T}_j$, $K_B$ being the Boltzmann constant. Similarly, the dimensionless time $t$ is defined as $t\coloneqq\frac{K}{\hbar}\bar{t}$, where $\hbar$ is the (reduced) Plank's constant, and $\bar{t}$ denotes the actual time. Hereafter, all quantities are taken as dimensionless throughout the paper.

In this case, the refrigerator is self-contained if the energy levels of the qubits satisfy the relation $E_3=E_2-E_1$.
 Further, we assume the baths are Markovian~\cite{10.1093/acprof:oso/9780199213900.001.0001}. Such baths cannot create local coherence and hence the local state of each qubit remains diagonal in the eigenbasis of the local Hamiltonian throughout the time evolution. This enables us to interpret the time-evolved single-qubit reduced states as thermal states and associate a value of temperature for each qubit at any time $t$.

Markovian baths are usually modelled~\cite{Mitchison_2015} as consisting of an infinite number of harmonic oscillators. Such baths are described by a local Hamiltonian of the form $H_{B_j}=\sum_{\textbf{k}}\nu_{j,\textbf{k}}b^{\dagger}_{j,\textbf{k}}b_{j,\textbf{k}}$, where the index $j$ refers that the bath is coupled to the $j$th qubit. The factor $\nu_{j,\textbf{k}}$ denotes the frequency corresponding to mode $\textbf{k}$, while the operators $b^{\dagger}_{j,\textbf{k}}$ and $b_{j,\textbf{k}}$ are the creation and annihilation operators of $\textbf{k}$th bosonic mode corresponding to the bath that is connected to the $j$th qubit. The local Hamiltonian of the three baths coupled to the three-qubit refrigerator system is then given by $H_B=\sum_{j=1}^3 H_{B_j}$. The system-bath interaction is described by the Hamiltonian $H_{SB}=\sum_{j=1}^3 \sigma^x_j\otimes\sum_{\textbf{k}}\left(\lambda_{j,\textbf{k}}b_{j,\textbf{k}}+\lambda^{*}_{j,\textbf{k}}b^{\dagger}_{j,\textbf{k}}\right)$, where $\sigma_j^x$ is the Pauli-$x$ matrix, which acts on the $j$th qubit. The constants $\lambda_{j,\textbf{k}}$ control the strength of coupling between the $j$th qubit and its corresponding bath. Thus, the total Hamiltonian of the composite system, consisting of the three-qubit refrigerator and the three baths, is given by $H_T=H_S+H_B+H_{SB}$. The bath coupled to $j$th qubit ($j=1,2,3$) is represented by the Ohmic spectral density function $J_j(\omega_j)=\alpha_j\omega_j \exp^{-\omega_j/\Omega}$. The set of frequencies $\{\omega_j\}$ correspond to the all possible differences in eigenvalues of the system Hamiltonian $H_S$. The system-bath interaction is therefore characterized by $\alpha_j$, which denotes the system-bath coupling strength, and a cutoff frequency $\Omega$. The cutoff frequency marks the regime of validity of the Markovian approximation \textemdash\ for a very large value of $\Omega$, the bath memory time (of order $\sim \Omega^{-1}$) is negligibly small and the baths are practically memoryless.  

We consider two scenarios based on the magnitude of the coupling strength $g$ between the qubits, compared to their energies. One is the strong coupling limit, where $g \geq E_j$, and the other is the weak coupling limit, obtained when $g << E_j$. The master equation derived from this model in the strong coupling limit is given by~\cite{Mitchison_2015}
\begin{equation}
\begin{aligned}
\frac{\partial\rho(t)}{\partial t} &= -i[H_S,\rho(t)] + \sum_{j,\omega_j} \gamma_j(\omega_j)\Biggl[\mathcal{L}^{\omega_j}_j\rho(t)\left(\mathcal{L}^{\omega_j}_j\right)^{\dagger} \\&- \frac{1}{2}\left\{\left(\mathcal{L}^{\omega_j}_j\right)^{\dagger}\mathcal{L}^{\omega_j}_j,\rho(t)\right\}\Biggr]. \label{scl}
\end{aligned}
\end{equation}
Where, $[A,B]\coloneqq AB-BA$ and $\{A,B\}\coloneqq AB+BA$. The operators $\left\{\mathcal{L}^{\omega_j}_j\right\}$ are the Lindblad operators that describe the various possible transitions of the system induced due to interaction with the baths. The transition rates $\gamma_j(\omega_j)$ corresponding to each  Lindblad operator are given as
\begin{equation}
\gamma_j(\omega_j) \coloneqq
\begin{cases}
J_j(\omega_j)\left[1 + n(\omega_j, \beta_j)\right], & (\omega_j > 0),\\
J_j(|\omega_j|)n(|\omega_j|, \beta_j), & (\omega_j < 0).
\end{cases}
\end{equation}
The factor $n(\omega_j, \beta_j)\coloneqq\left(\exp(\omega_j\beta_j)-1\right)^{-1}$ gives the mean occupation number of the Bose-Einstein distribution. The expressions of the Lindblad operators are given in Appendix~\ref{app}.

On the other hand, in the weak coupling limit, the dynamics of the qubits in the quantum refrigerator are given by~\cite{Mitchison_2015}
\begin{equation}
    \begin{aligned}
       \frac{\partial\rho(t)}{\partial t} &= -i[H_S,\rho(t)] + \sum_{j=1}^3 \biggl[\gamma_j(E_j)\mathcal{D}(\sigma^{-}_j)\rho(t)\\&+\gamma_j(-E_j)\mathcal{D}(\sigma^{+}_j)\rho(t)\biggr]. \label{wcl}
    \end{aligned}
\end{equation}
Where $\mathcal{D}(\mathcal{L})\rho\coloneqq\mathcal{L}\rho\mathcal{L}^{\dagger}-\frac{1}{2}\{\mathcal{L}^{\dagger}\mathcal{L},\rho\}$, and $\sigma_j^{\pm}\coloneqq\frac{1}{2}(\sigma_j^x\pm i\sigma_j^y)$ acts on the $j$th qubit. The operator $\sigma_j^y$ refers to the Pauli-$y$ matrix on the $j$th qubit.

\section{General framework for comparing physical quantities with finite precision}
\label{met}
In this section, we present a general framework for comparing the values of \emph{any} physical quantity pertaining to two - or more -
physical setups in the finite-precision scenario,
where the quantities are estimated using quantum metrology. In the latter part of this section, we demonstrate the application of this approach to quantum refrigeration.

In what follows, we present a general framework for comparing the values of any physical quantity pertaining to two physical setups, but it can be generalized to finitely many physical setups.
Let the physical quantity of interest be denoted by $Q$. Since exact estimation with zero error is unattainable, we assume that the estimated initial and final values of the quantity lie within finite ``patches'' $[Q_i^1, Q_i^2]$ and $[Q_f^1, Q_f^2]$, respectively, rather than being represented by single numbers. 
We say that the quantity \emph{decreases} (\emph{increases}) within finite precision if the lower (upper) bound of the initial interval, $Q_i^1$ ($Q_i^2$), is greater (less) than the upper (lower) bound of the final interval, $Q_f^2$ ($Q_f^1$). This condition ensures that every possible initial value within $[Q_i^1, Q_i^2]$ exceeds (falls below) all possible final values within $[Q_f^1, Q_f^2]$, thereby guaranteeing a definite decrease (increase) in the quantity despite finite estimation uncertainty. 
Thus, our formulation offers an unambiguous and operationally meaningful method for comparing physical quantities under realistic finite-precision conditions.

In parameter estimation theory, the standard deviation of an estimator is conventionally employed as a measure of estimation error. While this measure performs well when the estimator’s probability distribution is symmetric, it may fail to accurately capture the true uncertainty when the distribution is asymmetric - a situation that frequently arises in realistic metrological scenarios. To address this limitation, we introduce a universal quantifier of estimation error based on the percentiles of the estimator’s probability distribution, rather than its standard deviation. In what follows, we outline a step-by-step procedure for computing the percentiles of the estimator’s probability distribution for a general quantum parameter estimation task.

Let us denote the measurements performed in the eigenbasis of the SLD operator by ${Z_\theta}$, with ${p_Q}(\theta)$ representing the probabilities of obtaining the corresponding measurement outcomes $\theta$.

\begin{itemize}
    \item \textbf{Obtaining the probability distribution of the measurement outcomes:}  
    To determine the probability distribution of measurement outcomes, measurements are performed on the encoded quantum state in the eigenbasis of SLD operator corresponding to the parameter of interest.

    \item \textbf{Determining the estimator form:}  For our analysis, we consider the MVU estimator.
    By substituting the obtained probability distribution into the saturability condition of the (classical) Cram\'er-Rao bound, we derive the explicit form of the estimator $\hat{Q}(\theta)$.

    \item \textbf{Computing the moments of the estimator’s probability distribution:}  
    Using the form of the estimator, we compute the $n$th moment, $m_n$, of its probability distribution as
    \begin{equation*}
        m_n \coloneqq \sum_\theta \hat{Q}^n(\theta) \, p_Q(\theta); \quad n \in \mathbb{N}.
    \end{equation*}
  
    \item \textbf{Constructing the probability distribution of the estimator:}  
    From the first $M$ calculated moments, we construct the probability distribution function $f_{\hat{Q}}(Q)$ of the estimator $\hat{Q}(\theta)$ using the \textit{Maximum Entropy} (MaxEnt) principle~\cite{PhysRev.106.620,TAGLIANI1998157,10.1063/1.4819977}.  
    According to this principle, among all possible distributions consistent with the given $M$ moments, the least biased one is the distribution that maximizes the Shannon entropy,  $H(\{p_i\}) \coloneqq -\sum_i p_i \log_2 p_i$ where $\{p_i\}$ denotes the distribution.

    \item \textbf{Determining the percentiles of the estimator:}  
   The cumulative distribution function (CDF) $\mathbb{F}_{\hat{Q}}(Q)$ associated with $f_{\hat{Q}}(Q)$ is given by
    \begin{equation*}
        \mathbb{F}_{\hat{Q}}(\bar{Q}) = \int_{-\infty}^{\bar{Q}} f_{\hat{Q}}(Q) \, dQ.
    \end{equation*}
    The $i$th percentile, $\bar{Q}_i$, corresponds to the value of $\bar{Q}$ satisfying
    \begin{equation}
        \mathbb{F}_{\hat{Q}}(\bar{Q}_i) = \int_{-\infty}^{\bar{Q}_i} f_{\hat{Q}}(Q) \, dT = \frac{i}{100},
    \end{equation}
    where $i \in [1,100]$.

    \item \textbf{Checking the convergence of percentiles:}  
    To ensure the reliability of the estimated percentiles, the entire procedure is repeated using the next $(M+1)$ moments. The convergence of percentiles is verified by checking whether the percentile values remain within a predefined tolerance compared to those obtained using the first $M$ moments. This process is iterated for successive moments ($M+1$, $M+2$, etc.) until convergence is achieved.
\end{itemize}

\section{
Finite-precision cooling in quantum processes}
\label{res}
In this section, we apply our general framework for comparing physical quantities to study cooling in quantum systems under finite-precision conditions. We specifically focus on the three-qubit self-contained quantum refrigerator governed by Markovian dynamics, as discussed in Sec.~\ref{qref}. 
The section is twofold: in the first part, we derive the forms of the estimator corresponding to the temperature of the cold qubit in a quantum refrigerator; in the second part, we demonstrate the occurrence of finite-precision cooling using the percentile-based framework in this model.

We begin by deriving the form of the MVU estimator for the temperature of the cold qubit. Since the refrigerator dynamics are Markovian~\cite{Mitchison_2015}, the state of the cold qubit at any instant can be described by a thermal (Gibbs) state
\begin{equation*}
    \tilde{\tau}_1 = \frac{\exp(-\tilde{\beta}_1 H_1)}{\Tr[\exp(-\tilde{\beta}_1 H_1)]},
\end{equation*}
where $H_1 = E_1 \sigma_1^z / 2$ is the local Hamiltonian and $\tilde{\beta}_1 = 1 / \tilde{T}_1$, with $\tilde{T}_1$ being the instantaneous temperature of the qubit. 
In the energy eigenbasis, this can be written as 
\begin{equation*}
    \tilde{\tau}_1 = r_1 \ket{0}\bra{0} + (1 - r_1) \ket{1}\bra{1},
\end{equation*} 
 where 
\begin{equation*}
    r_1 \coloneqq \frac{e^{\tilde{\beta}_1 E_1 / 2}}{e^{\tilde{\beta}_1 E_1 / 2} + e^{-\tilde{\beta}_1 E_1 / 2}}
\end{equation*}
is the ground-state population.
Using Eq.~\eqref{sld}, the SLD operator is determined as
\begin{equation}
\label{eq-SLD-cooling}
    \Lambda_T = \frac{\dot{r}_1}{r_1}\ket{0}\bra{0} - \frac{\dot{r}_1}{1 - r_1}\ket{1}\bra{1},
\end{equation}
where $\dot{r}_1 = \frac{dr_1}{d\tilde{T}_1}$. 
Thus, the SLD eigenbasis coincides with the energy eigenbasis $\{\ket{0}, \ket{1}\}$.

 Let the outcomes of the energy measurement be denoted by $\{\mathcal{E}_i\}$. 
The corresponding probability distribution for the $i$th outcome is 
$p_{\tilde{T}_1}(\mathcal{E}_i) = \{\Tr(\ket{0}\bra{0}\tilde{\tau}_1), \Tr(\ket{1}\bra{1}\tilde{\tau}_1)\} = \{r_1, 1 - r_1\}$. 
Equivalently, it can be expressed in terms of $\tilde{\beta}_1$ and the energy eigenvalues $\{\mathcal{E}_i\}$ as 
\begin{equation*}
    p_{\tilde{T}_1}(\mathcal{E}_i) = \frac{1}{Z_1} e^{-\tilde{\beta}_1 \mathcal{E}_i},
\end{equation*}
where $Z_1 \coloneqq \sum_i e^{-\tilde{\beta}_1 \mathcal{E}_i}$ is the partition function of the cold qubit.
Using this expression, the left-hand side of Eq.~\eqref{satur} becomes
\begin{align}
    \frac{d \ln p_{\tilde{T}_1}(\mathcal{E}_i)}{d\tilde{T}_1} 
    &= \frac{d}{d\tilde{T}_1}\left[-\frac{\mathcal{E}_i}{\tilde{T}_1} - \ln Z_1\right] \nonumber\\
    &= \frac{\mathcal{E}_i - \langle H_1 \rangle}{\tilde{T}_1^2}, \label{p_T}
\end{align}
where we have used the thermodynamic relation 
$\langle H_1 \rangle = \tilde{T}_1^2 \frac{d}{d\tilde{T}_1}\ln Z_1$ 
for the average energy of a canonical ensemble. 
Substituting Eq.~\eqref{p_T} into Eq.~\eqref{satur} and using $F_C = F_Q$, we obtain the following form of the MVU estimator for the cold qubit at any temperature during its evolution:
\begin{equation}
    \hat{\tilde{T}}_1(\mathcal{E}_i) 
    = \frac{1}{F_Q}\frac{d \ln p_{\tilde{T}_1}(\mathcal{E}_i)}{d\tilde{T}_1} + \tilde{T}_1 
    = \frac{\left(\mathcal{E}_i - \langle H_1 \rangle\right)}{F_Q \tilde{T}_1^2} + \tilde{T}_1. 
    \label{estimator}
\end{equation}

{Since, in our case, the MVU estimator is a function of $F_Q$, we obtain the MVU estimator by substituting the form of the SLD operator (Eq.~\eqref{eq-SLD-cooling}) in the expression of $F_Q$.}

To verify finite-precision cooling in quantum refrigerators, we use the form of the MVU estimator in
Eq.~\eqref{estimator} 
to compute the converged percentiles for both the initial and final temperatures of the cold qubit, following the procedure outlined in Sec.~\ref{met}. 
Since the Shannon entropy $H(\{p_i\})$ is a concave function, maximizing it is equivalent to minimizing $-H(\{p_i\})$, which is convex. Hence, the MaxEnt principle can be reformulated as a convex optimization problem. We have used the CVXPY package~\cite{diamond2016cvxpy,agrawal2018rewriting} of Python and the solver MOSEK~\cite{mosek} to numerically perform the optimization. In the procedure of computing the percentiles, one needs to solve this convex optimization problem. We have used the CVXPY package~\cite{diamond2016cvxpy} to numerically perform the optimization. 
Finite-precision cooling is said to occur at the $i$-th percentile if the cooling magnitude
$\Delta\bar{T}_i \coloneqq \bar{T}_i-\bar{T}^f_{100-i} > 0$,
where $\bar{T}_i$ and $\bar{T}^f_{100-i}$ denote the $i$-th percentile of the initial and $(100-i)$-th percentile of the final temperature distributions of the cold qubit, respectively. Larger values of $\Delta\bar{T}_i$ correspond to stronger cooling under finite-precision conditions. The occurrence of cooling thus depends on the particular percentile used for comparison. If no percentile $i < 50$ satisfies the finite-precision cooling condition, we conclude that cooling does not occur within the limits of finite precision.

The three-qubit self-contained refrigerator operates under Markovian dynamics governed by the Lindblad equations, Eq.~\eqref{scl} and Eq.~\eqref{wcl}, corresponding to the strong and weak inter-qubit interaction coupling limits, respectively. For given system and bath parameters, we numerically solve these equations to obtain the cold-qubit temperature at each instant of evolution. For both regimes, we study a broad range of final temperatures covering transient and steady-state behavior. Subsection~\ref{sc} presents examples of steady-state and transient finite-precision cooling in the strong-inter-qubit interaction coupling limit, while subsection~\ref{wc} discusses analogous results for the weak inter-qubit interaction coupling limit. In all cases, the dynamics are evolved over the interval $0 \leq t \leq 10^5$, with the system reaching steady state well before $t = 10^5$. Moreover, in both the coupling limits, we have obtained the convergence (up to the second decimal place) of all the percentiles, only considering the first two moments and the first three moments of the estimator.

\begin{figure*}[t]
    \centering
    \begin{minipage}[t]{0.45\textwidth}
        \centering
        \includegraphics[trim=0cm 0.0cm 0.0cm 0, clip,width=9cm, height=5.9cm]{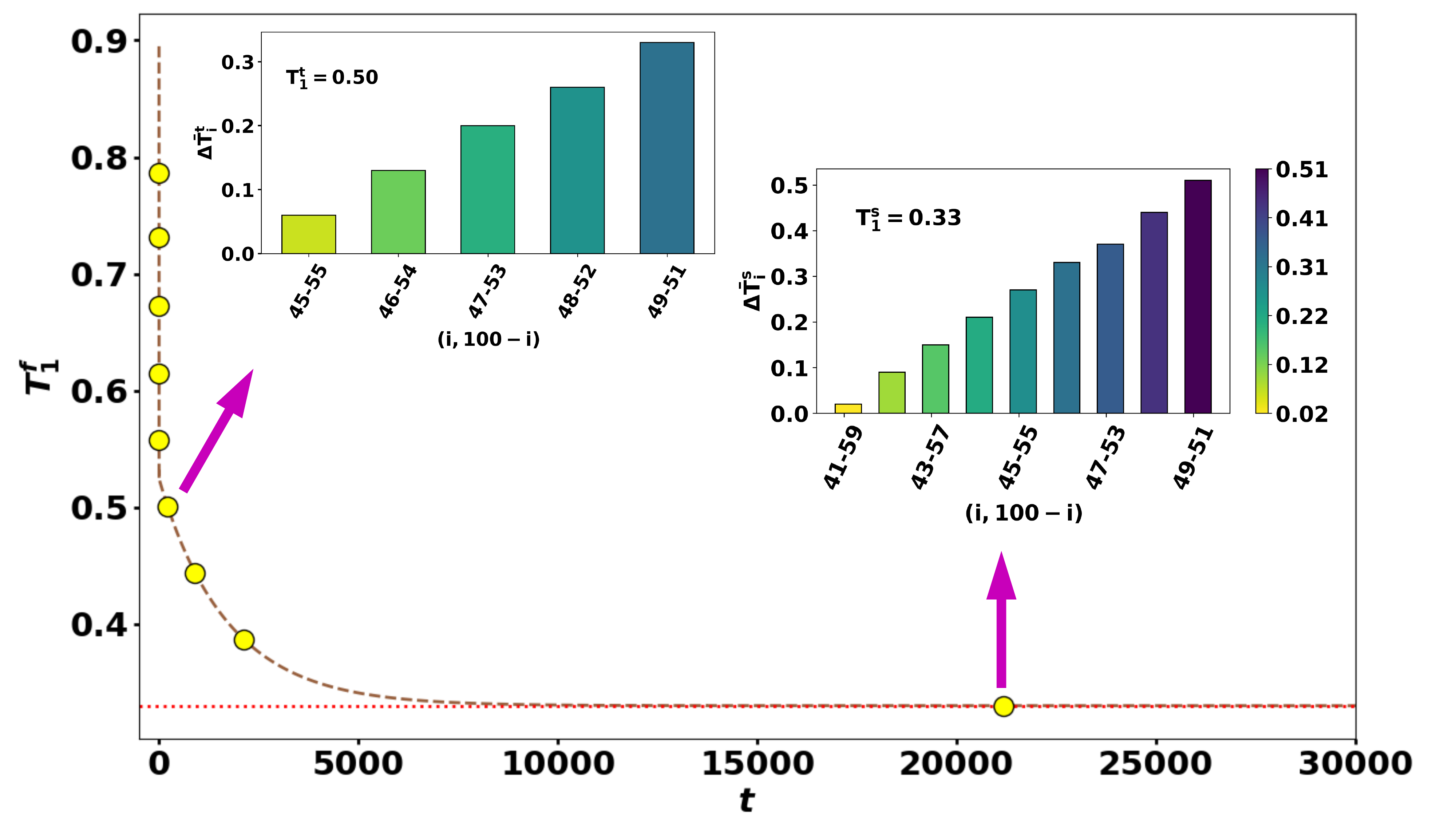}\\
        \textbf{(a)}
    \end{minipage}
    \hfill
    \begin{minipage}[t]{0.45\textwidth}
        \centering
        \includegraphics[trim=0cm 0.0cm 0.0cm 0, clip, width=8cm, height=6.6cm]{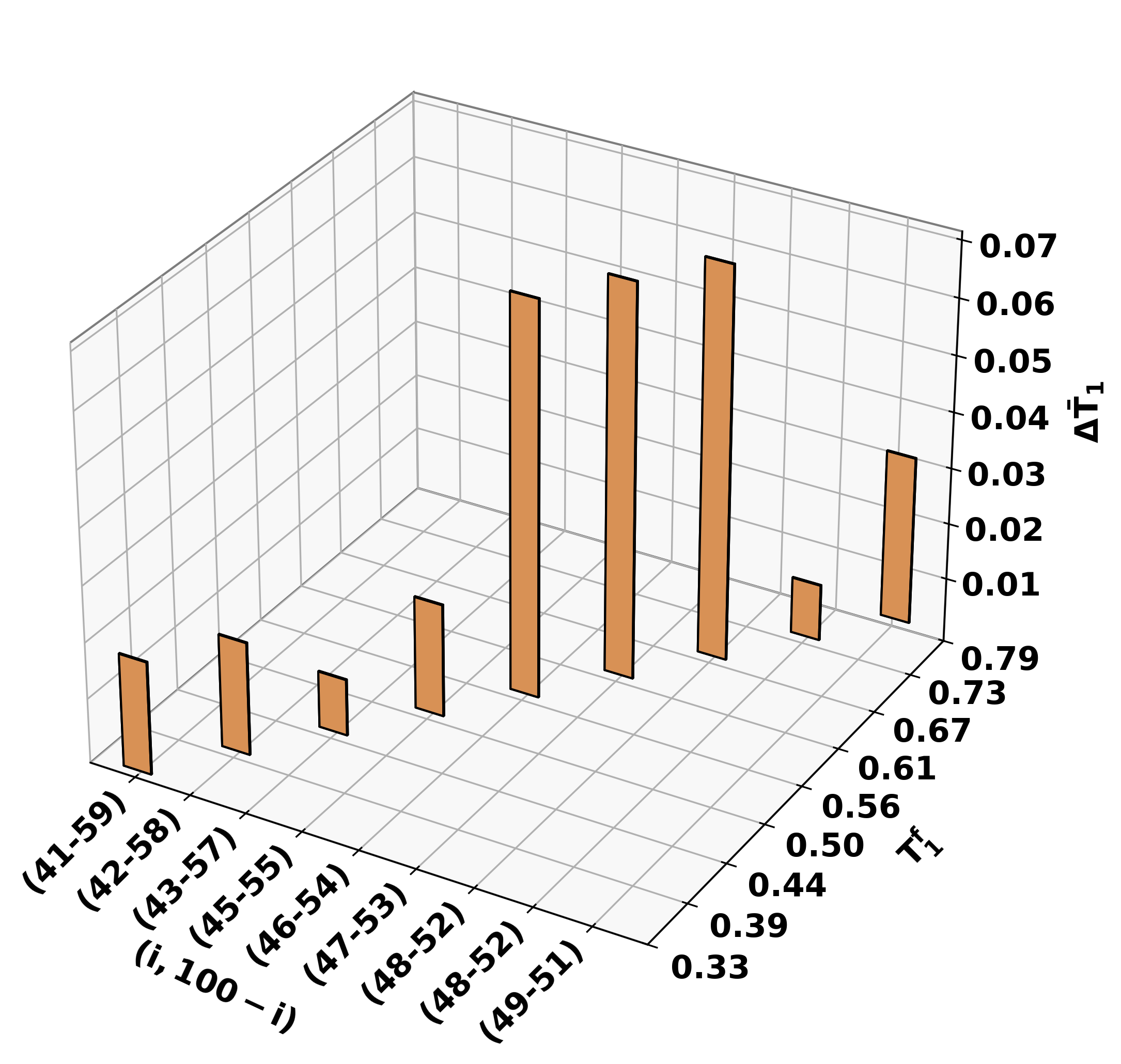}\\
        \textbf{(b)}
    \end{minipage}
    \caption{\textbf{Finite-precision cooling of the cold qubit in a 3-qubit quantum absorption refrigerator under strong-coupling limit.} \textbf{(a)} The temperature $T^f_1(t)$ of the qubit (brown dashed line) is plotted as a function of time $t$, with the steady-state temperature $T^s_1=0.33$ indicated by the red dotted line. Time is shown on the horizontal axis, and temperature on the vertical axis. To examine cooling in the finite-precision sense, we select $9$ representative points along $T^f_1(t)$, spanning from the transient to steady-state regimes. The superscripts $f$ and $s$ denote transient and steady-state values, respectively, as illustrated in the insets. The points are marked in yellow circles and are approximately equally spaced with an interval of $0.05$. The first $8$ points correspond to the transient regime, while the final point represents the steady state. The insets highlight cooling at two points (pink arrows): a transient value $T^t_1(t)=0.50$ (left) and the steady-state temperature $T^s_1=0.33$ (right). For each inset, the horizontal axis shows the percentile range over which cooling occurs, and the vertical axis plots the cooling magnitude $\Delta\bar{T}_i$ for each percentile $i$. Percentile ranges in the inset for $T^s_1$ are labeled alternately along the horizontal axis for clarity. \textbf{(b)} The $3$D plot presents the percentiles $(i,100-i)$, where $i$ is a fixed, positive integer denoting the percentile for which the finite-precision cooling occurs for the first time for each of the $9$ points from panel (a), along with the points and the cooling magnitudes. The horizontal plane shows the  percentiles and corresponding data points (rounded up to $2$ decimal places), while the vertical axis represents the cooling magnitudes $\Delta\bar{T}_1$ ( yellow bars). All parameters plotted are dimensionless.}
    \label{fig:1} 
\end{figure*}

    \begin{figure*}[t]
    \centering
    \begin{minipage}[t]{0.45\textwidth}
        \centering
        \includegraphics[trim=0cm 0.0cm 0.0cm 0, clip,width=9cm, height=5.9cm]{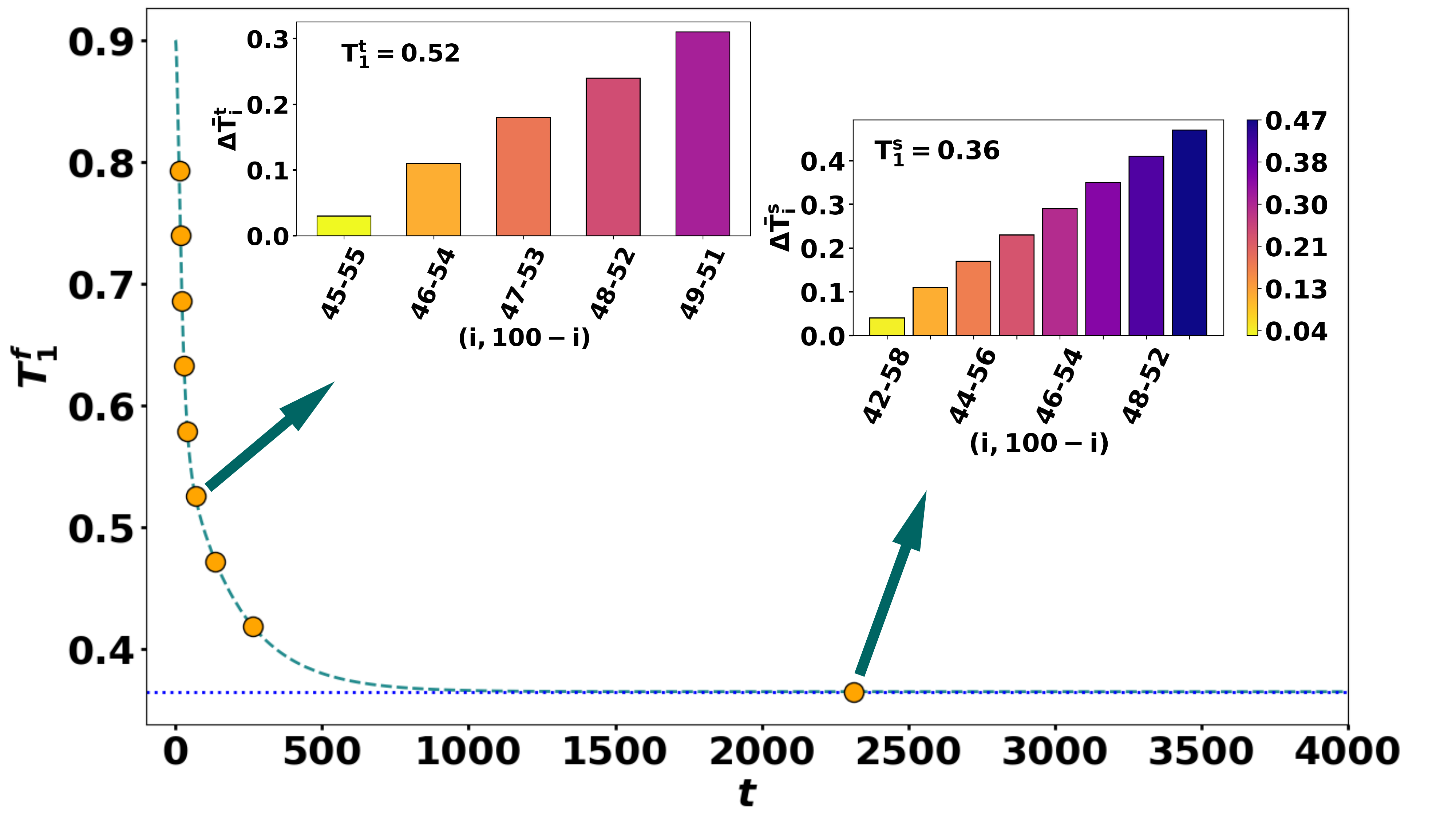}\\
        \textbf{(a)}
    \end{minipage}
    \hfill
    \begin{minipage}[t]{0.45\textwidth}
        \centering
        \includegraphics[trim=0cm 0.0cm 0.0cm 0, clip, width=8cm, height=6.6cm]{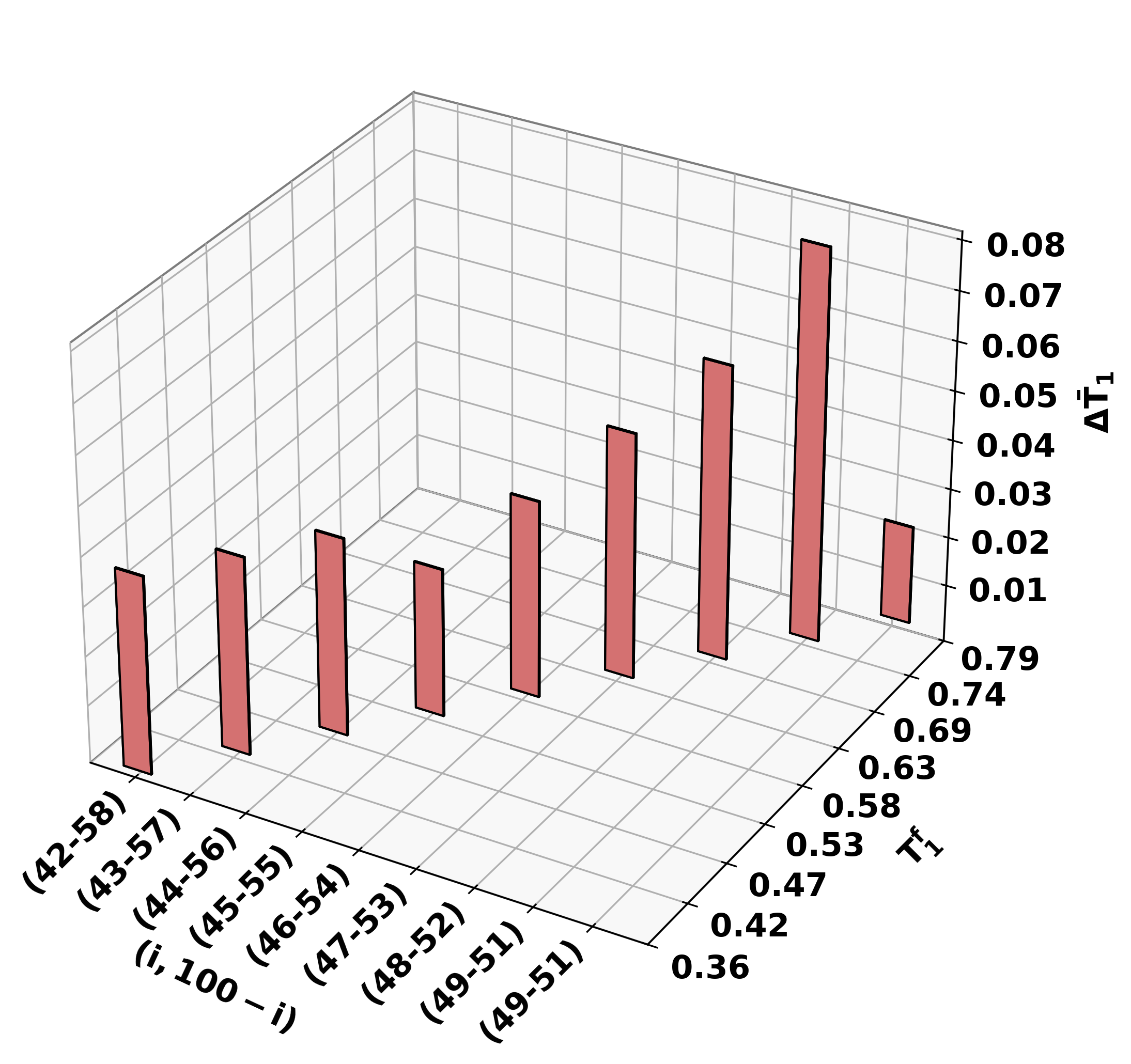}\\
        \textbf{(b)}
    \end{minipage}   \caption{\textbf{Illustration of finite-precision cooling of the cold qubit in a 3-qubit quantum absorption refrigerator in the weak-coupling limit.} \textbf{(a)} Variation of temperature $T^f_1(t)$ of the qubit (green dashed line), with time $t$. The time is plotted along the horizontal axis, whereas the temperature is plotted along the vertical axis. The blue dotted line is the steady-state temperature of the qubit, $T^s_1=0.36$. To explicitly check for cooling using percentiles, we consider $9$ data points from the curve of the qubit temperature $T^f_1(t)$, across the transient and the steady-state regions. Here, the superscript $f$ is labelled as $t$ (transient) and $s$ (steady state), as illustrated in the inset plots. The points, marked by orange circles, are evenly spaced with an interval of approximately $0.05$. The first $8$ points correspond to the transient regime, while the final point represents the steady-state temperature. The insets highlight the occurrence of cooling for two points (indicated by green arrows): the steady state value $T^s_1=0.36$ (right), and a transient value $T^t_1(t)=0.52$ (left). In each of the inset plots, the percentile range across which the cooling occurs is plotted along the horizontal axis, while the corresponding cooling magnitudes $\Delta\bar{T}_i$, for each value $i$ within the percentile range is plotted along the vertical axis. For better readability, percentile ranges are labelled alternately along the horizontal axis in the inset for $T^s_1$.  \textbf{(b)} The $3$D plot depicts the percentiles $(i,100-i)$, where $i$ is a fixed positive integer indicating the point at which cooling, in the finite-precision sense, is first detected for each of the $9$ data points considered in panel (a), along with the data points and the corresponding cooling magnitudes. The percentiles and the data points, rounded to $2$ decimal places, are plotted in the horizontal plane, and the corresponding cooling magnitudes $\Delta\bar{T}_1$ (shown by light red bars) are plotted along the vertical axis. All quantities appearing in the plots are dimensionless.}
    \label{fig:2}
\end{figure*}

\subsection{Strong inter-qubit interaction coupling limit} 
\label{sc}

The strong coupling limit is obtained when the inter-qubit coupling strength of the three-qubit refrigerator system is comparable to the qubit energies.

Eq.~\eqref{scl} is numerically solved by considering the following choice of system and bath parameters. We choose 
$E_1=1.0, E_2=10.0$ for the cold and the work qubit respectively. To make the refrigerator self-contained, we set $E_3=E_2-E_1=9.0$ for the hot qubit. The initial temperatures of the qubits are taken to be $T_1=T_2=0.9, \;T_3=100.0$. The inter-qubit coupling strength is taken as $g=0.8$. The system-bath coupling strengths are taken to be $\alpha_1=10^{-4},\alpha_2=10^{-4},\alpha_3=10^{-2}$. The cutoff frequencies of the baths are  set as $\Omega_1=\Omega_2=\Omega_3=10^4$.

The results are presented in Fig.~\ref{fig:1}. Panel (a) shows the evolution of the cold qubit temperature $T^f_1(t)$ over time $t$. We plot the time on the horizontal axis and the temperature on the vertical axis, restricting the time range up to $t=30000$. The qubit attains the steady state at a time $t =17025.10$, with a numerical precision up to second decimal place. 

The temperature dynamics is illustrated by the brown dashed curve, with the steady-state value $T^s_1=0.33$ indicated by a red dotted line. To analyze finite-precision cooling, we computed percentiles for $9$ representative points spanning $T^f_1(t) \in [0.9,0.33]$. These points, marked by yellow circles, are approximately equally spaced with an interval of 0.05. Eight points lie within the transient regime, while the final point corresponds to the steady-state temperature. This ensures that cooling is obtained across the entire dynamics.

The insets in panel (a) show the detailed cooling profiles for a transient value $T^t_1=0.50$ (left) and the steady-state temperature $T^s_1$ (right). For each inset, the horizontal axis indicates the percentile range over which cooling occurs, while the vertical axis shows the corresponding cooling magnitude $\Delta\bar{T}_i$. As observed, the cooling magnitude increases as the percentile range is narrowed down (corresponding to increasing $i$).

Panel (b) summarizes the first occurrence of finite-precision cooling for each of the $9$ points in panel (a). We plot the percentiles $(i,100-i)$, along with the data points and the corresponding cooling magnitudes $\Delta\bar{T}_i$. The horizontal plane contains the percentiles and the data points rounded to $2$ decimal places, whereas the vertical axis depicts the cooling magnitudes (shown by yellow bars). Here, $i$ is a fixed positive integer ($<50$) that identifies the percentile at which cooling is first detected.

\subsection{Weak inter-qubit interaction coupling limit}
\label{wc}
In this subsection, we show finite precision cooling for a given range of final temperature in the weak-coupling limit, obtained for coupling strengths much smaller compared to the qubit energies.
A numerical solution to Eq.~\eqref{wcl} is obtained for the following choice of system and bath parameters. The qubit energies and their initial temperatures, as well as the cutoff frequencies of the baths are taken to be the same as~\ref{sc}. The inter-qubit coupling strength is set as $g=0.05$. The system-bath coupling strengths are  taken to be $\alpha_1=\alpha_2=\alpha_3=10^{-3}$.

The results are demonstrated in Fig.~\ref{fig:2}. The variation of the temperature of the cold qubit $T^f_1(t)$ with the time $t$ of the dynamics is depicted in panel (a). The horizontal axis represents time, while the vertical axis represents temperature. The range of the time-axis is restricted up to $t=4000$, where the steady state is reached at $t=2313.53$, accurate to the second decimal place.

The behavior of $T^f_1(t)$ is represented by the green dashed curve, while the blue dotted horizontal line shows the steady-state temperature $T^s_1=0.36$. We have explicitly computed the percentiles and observed cooling for $9$ data points across the temperature range $T^f_1(t) \in [0.9,0.36]$. The points are shown by orange circles and are equispaced with spacing of approximately 0.05. Among them, eight points correspond to the transient regime and the last point in right side corresponds to the steady-state temperature denoted as $T^s_1$ . We have thus observed transient cooling as well as steady-state cooling, in the finite-precision sense. The two plots in the inset depict the complete cooling profile for $T^s_1$ (plot on the right) and a transient temperature $T^t_1=0.52$ (plot on the left). In each inset, we plot the percentile range for which cooling occurs and the corresponding cooling magnitudes $\Delta\bar{T}_i$, where $i$ varies over the cooling range. Similar to subsec.~\ref{sc}, it is seen from the insets that as the percentile range becomes narrower ($i$ increases), the cooling magnitude increases. In panel (b), we plot the percentiles $(i,100-i)$ for which the cooling (in the finite-precision sense) is obtained for the first time for each of the $9$ data points in panel (a), along with the data points and the associated cooling magnitudes $\Delta\bar{T}_i$. Here, $i$ is a fixed positive integer ($<50$) labeling the percentile corresponding to the first detection of finite-precision cooling for each data point. The cooling magnitudes are depicted as light red bars. The percentiles and the data points, rounded to $2$ decimal places, are shown across the horizontal plane, and the vertical axis plots the cooling magnitudes.

\section{Conclusion}
\label{con}
In this work, we focused on comparing the values of a physical quantity corresponding to two different situations within the paradigm of quantum metrology.
Since metrological estimation naturally yields a
``patch” on the real line rather than a single number
representing estimation error, a meaningful framework must account for the comparison of the values of a physical quantity within the limits of finite precision.
Building on this idea, we developed an approach that allows the comparison of two physical quantities within the limits of finite precision. As an application, we introduced the concept of finite-precision cooling, a notion of cooling that incorporates the constraints imposed by finite-precision estimation.

Typically, the precision of a parameter estimate is quantified using the standard deviation of the estimator. However, when the estimator’s probability distribution is asymmetric, the standard deviation may fail to capture the actual uncertainty. To address this, we proposed a universal approach based on the percentiles of the estimator’s probability distribution, providing a robust quantification of uncertainty applicable to both symmetric and asymmetric distributions.
We applied this approach to examine cooling in a three-qubit quantum absorption refrigerator governed by Markovian dynamics. The temperatures of the cold qubit in both the initial and time-evolved states were estimated using the minimum-variance-unbiased estimator, with measurements performed in the eigenbasis of the symmetric logarithmic derivative operator. Our analysis considered both strong- and weak-coupling limits of the inter-qubit interaction, with numerical examples spanning the transient to steady-state regimes. In all cases, cooling was observed within the limits of finite precision.

Overall, our results provide a general methodology for comparing physical quantities under realistic, finite-precision constraints, extending the scope of quantum metrology to comparative physical problems and offering new insights into quantum processes such as cooling of quantum systems.

\section{Acknowledgments}
We acknowledge the use of Armadillo, QIClib (\url{https://titaschanda.github.io/QIClib}), CVXPY and MOSEK. 
PG acknowledges support from the ‘INFOSYS scholarship for senior students’ at Harish-Chandra Research Institute, India.
\bibliography{ref}

\begin{thebibliography}{76}%
\makeatletter
\providecommand \@ifxundefined [1]{%
 \@ifx{#1\undefined}
}%
\providecommand \@ifnum [1]{%
 \ifnum #1\expandafter \@firstoftwo
 \else \expandafter \@secondoftwo
 \fi
}%
\providecommand \@ifx [1]{%
 \ifx #1\expandafter \@firstoftwo
 \else \expandafter \@secondoftwo
 \fi
}%
\providecommand \natexlab [1]{#1}%
\providecommand \enquote  [1]{``#1''}%
\providecommand \bibnamefont  [1]{#1}%
\providecommand \bibfnamefont [1]{#1}%
\providecommand \citenamefont [1]{#1}%
\providecommand \href@noop [0]{\@secondoftwo}%
\providecommand \href [0]{\begingroup \@sanitize@url \@href}%
\providecommand \@href[1]{\@@startlink{#1}\@@href}%
\providecommand \@@href[1]{\endgroup#1\@@endlink}%
\providecommand \@sanitize@url [0]{\catcode `\\12\catcode `\$12\catcode `\&12\catcode `\#12\catcode `\^12\catcode `\_12\catcode `\%12\relax}%
\providecommand \@@startlink[1]{}%
\providecommand \@@endlink[0]{}%
\providecommand \url  [0]{\begingroup\@sanitize@url \@url }%
\providecommand \@url [1]{\endgroup\@href {#1}{\urlprefix }}%
\providecommand \urlprefix  [0]{URL }%
\providecommand \Eprint [0]{\href }%
\providecommand \doibase [0]{https://doi.org/}%
\providecommand \selectlanguage [0]{\@gobble}%
\providecommand \bibinfo  [0]{\@secondoftwo}%
\providecommand \bibfield  [0]{\@secondoftwo}%
\providecommand \translation [1]{[#1]}%
\providecommand \BibitemOpen [0]{}%
\providecommand \bibitemStop [0]{}%
\providecommand \bibitemNoStop [0]{.\EOS\space}%
\providecommand \EOS [0]{\spacefactor3000\relax}%
\providecommand \BibitemShut  [1]{\csname bibitem#1\endcsname}%
\let\auto@bib@innerbib\@empty
\bibitem [{\citenamefont {Helstrom}(1969)}]{Helstrom1969}%
  \BibitemOpen
  \bibfield  {author} {\bibinfo {author} {\bibfnamefont {C.~W.}\ \bibnamefont {Helstrom}},\ }\bibfield  {title} {\bibinfo {title} {Quantum detection and estimation theory},\ }\href {https://doi.org/10.1007/BF01007479} {\bibfield  {journal} {\bibinfo  {journal} {J. Stat. Phys.}\ }\textbf {\bibinfo {volume} {1}},\ \bibinfo {pages} {231} (\bibinfo {year} {1969})}\BibitemShut {NoStop}%
\bibitem [{\citenamefont {Holevo}(1973)}]{HOLEVO1973337}%
  \BibitemOpen
  \bibfield  {author} {\bibinfo {author} {\bibfnamefont {A.~S.}\ \bibnamefont {Holevo}},\ }\bibfield  {title} {\bibinfo {title} {Statistical decision theory for quantum systems},\ }\href {https://doi.org/https://doi.org/10.1016/0047-259X(73)90028-6} {\bibfield  {journal} {\bibinfo  {journal} {JMVA}\ }\textbf {\bibinfo {volume} {3}},\ \bibinfo {pages} {337} (\bibinfo {year} {1973})}\BibitemShut {NoStop}%
\bibitem [{\citenamefont {Braunstein}\ and\ \citenamefont {Caves}(1994)}]{PhysRevLett.72.3439}%
  \BibitemOpen
  \bibfield  {author} {\bibinfo {author} {\bibfnamefont {S.~L.}\ \bibnamefont {Braunstein}}\ and\ \bibinfo {author} {\bibfnamefont {C.~M.}\ \bibnamefont {Caves}},\ }\bibfield  {title} {\bibinfo {title} {Statistical distance and the geometry of quantum states},\ }\href {https://doi.org/10.1103/PhysRevLett.72.3439} {\bibfield  {journal} {\bibinfo  {journal} {Phys. Rev. Lett.}\ }\textbf {\bibinfo {volume} {72}},\ \bibinfo {pages} {3439} (\bibinfo {year} {1994})}\BibitemShut {NoStop}%
\bibitem [{\citenamefont {Giovannetti}\ \emph {et~al.}(2004)\citenamefont {Giovannetti}, \citenamefont {Lloyd},\ and\ \citenamefont {Maccone}}]{doi:10.1126/science.1104149}%
  \BibitemOpen
  \bibfield  {author} {\bibinfo {author} {\bibfnamefont {V.}~\bibnamefont {Giovannetti}}, \bibinfo {author} {\bibfnamefont {S.}~\bibnamefont {Lloyd}},\ and\ \bibinfo {author} {\bibfnamefont {L.}~\bibnamefont {Maccone}},\ }\bibfield  {title} {\bibinfo {title} {Quantum-enhanced measurements: Beating the standard quantum limit},\ }\href {https://doi.org/10.1126/science.1104149} {\bibfield  {journal} {\bibinfo  {journal} {Science}\ }\textbf {\bibinfo {volume} {306}},\ \bibinfo {pages} {1330} (\bibinfo {year} {2004})}\BibitemShut {NoStop}%
\bibitem [{\citenamefont {Cappellaro}\ \emph {et~al.}(2005)\citenamefont {Cappellaro}, \citenamefont {Emerson}, \citenamefont {Boulant}, \citenamefont {Ramanathan}, \citenamefont {Lloyd},\ and\ \citenamefont {Cory}}]{PhysRevLett.94.020502}%
  \BibitemOpen
  \bibfield  {author} {\bibinfo {author} {\bibfnamefont {P.}~\bibnamefont {Cappellaro}}, \bibinfo {author} {\bibfnamefont {J.}~\bibnamefont {Emerson}}, \bibinfo {author} {\bibfnamefont {N.}~\bibnamefont {Boulant}}, \bibinfo {author} {\bibfnamefont {C.}~\bibnamefont {Ramanathan}}, \bibinfo {author} {\bibfnamefont {S.}~\bibnamefont {Lloyd}},\ and\ \bibinfo {author} {\bibfnamefont {D.~G.}\ \bibnamefont {Cory}},\ }\bibfield  {title} {\bibinfo {title} {Entanglement assisted metrology},\ }\href {https://doi.org/10.1103/PhysRevLett.94.020502} {\bibfield  {journal} {\bibinfo  {journal} {Phys. Rev. Lett.}\ }\textbf {\bibinfo {volume} {94}},\ \bibinfo {pages} {020502} (\bibinfo {year} {2005})}\BibitemShut {NoStop}%
\bibitem [{\citenamefont {Giovannetti}\ \emph {et~al.}(2006)\citenamefont {Giovannetti}, \citenamefont {Lloyd},\ and\ \citenamefont {Maccone}}]{PhysRevLett.96.010401}%
  \BibitemOpen
  \bibfield  {author} {\bibinfo {author} {\bibfnamefont {V.}~\bibnamefont {Giovannetti}}, \bibinfo {author} {\bibfnamefont {S.}~\bibnamefont {Lloyd}},\ and\ \bibinfo {author} {\bibfnamefont {L.}~\bibnamefont {Maccone}},\ }\bibfield  {title} {\bibinfo {title} {Quantum metrology},\ }\href {https://doi.org/10.1103/PhysRevLett.96.010401} {\bibfield  {journal} {\bibinfo  {journal} {Phys. Rev. Lett.}\ }\textbf {\bibinfo {volume} {96}},\ \bibinfo {pages} {010401} (\bibinfo {year} {2006})}\BibitemShut {NoStop}%
\bibitem [{\citenamefont {Giovannetti}\ \emph {et~al.}(2011)\citenamefont {Giovannetti}, \citenamefont {Lloyd},\ and\ \citenamefont {Maccone}}]{Giovannetti2011}%
  \BibitemOpen
  \bibfield  {author} {\bibinfo {author} {\bibfnamefont {V.}~\bibnamefont {Giovannetti}}, \bibinfo {author} {\bibfnamefont {S.}~\bibnamefont {Lloyd}},\ and\ \bibinfo {author} {\bibfnamefont {L.}~\bibnamefont {Maccone}},\ }\bibfield  {title} {\bibinfo {title} {Advances in quantum metrology},\ }\href {https://doi.org/10.1038/nphoton.2011.35} {\bibfield  {journal} {\bibinfo  {journal} {Nat. Photon.}\ }\textbf {\bibinfo {volume} {5}},\ \bibinfo {pages} {222} (\bibinfo {year} {2011})}\BibitemShut {NoStop}%
\bibitem [{\citenamefont {Holevo}(2011)}]{holevo}%
  \BibitemOpen
  \bibfield  {author} {\bibinfo {author} {\bibfnamefont {A.~S.}\ \bibnamefont {Holevo}},\ }\href@noop {} {\emph {\bibinfo {title} {Probabilistic and statistical aspects of quantum theory, Vol. 1}}}\ (\bibinfo  {publisher} {Springer Science \& Business Media},\ \bibinfo {year} {2011})\BibitemShut {NoStop}%
\bibitem [{\citenamefont {Degen}\ \emph {et~al.}(2017)\citenamefont {Degen}, \citenamefont {Reinhard},\ and\ \citenamefont {Cappellaro}}]{RevModPhys.89.035002}%
  \BibitemOpen
  \bibfield  {author} {\bibinfo {author} {\bibfnamefont {C.~L.}\ \bibnamefont {Degen}}, \bibinfo {author} {\bibfnamefont {F.}~\bibnamefont {Reinhard}},\ and\ \bibinfo {author} {\bibfnamefont {P.}~\bibnamefont {Cappellaro}},\ }\bibfield  {title} {\bibinfo {title} {Quantum sensing},\ }\href {https://doi.org/10.1103/RevModPhys.89.035002} {\bibfield  {journal} {\bibinfo  {journal} {Rev. Mod. Phys.}\ }\textbf {\bibinfo {volume} {89}},\ \bibinfo {pages} {035002} (\bibinfo {year} {2017})}\BibitemShut {NoStop}%
\bibitem [{\citenamefont {Pezz\`e}\ \emph {et~al.}(2018)\citenamefont {Pezz\`e}, \citenamefont {Smerzi}, \citenamefont {Oberthaler}, \citenamefont {Schmied},\ and\ \citenamefont {Treutlein}}]{RevModPhys.90.035005}%
  \BibitemOpen
  \bibfield  {author} {\bibinfo {author} {\bibfnamefont {L.}~\bibnamefont {Pezz\`e}}, \bibinfo {author} {\bibfnamefont {A.}~\bibnamefont {Smerzi}}, \bibinfo {author} {\bibfnamefont {M.~K.}\ \bibnamefont {Oberthaler}}, \bibinfo {author} {\bibfnamefont {R.}~\bibnamefont {Schmied}},\ and\ \bibinfo {author} {\bibfnamefont {P.}~\bibnamefont {Treutlein}},\ }\bibfield  {title} {\bibinfo {title} {Quantum metrology with nonclassical states of atomic ensembles},\ }\href {https://doi.org/10.1103/RevModPhys.90.035005} {\bibfield  {journal} {\bibinfo  {journal} {Rev. Mod. Phys.}\ }\textbf {\bibinfo {volume} {90}},\ \bibinfo {pages} {035005} (\bibinfo {year} {2018})}\BibitemShut {NoStop}%
\bibitem [{\citenamefont {Braun}\ \emph {et~al.}(2018)\citenamefont {Braun}, \citenamefont {Adesso}, \citenamefont {Benatti}, \citenamefont {Floreanini}, \citenamefont {Marzolino}, \citenamefont {Mitchell},\ and\ \citenamefont {Pirandola}}]{RevModPhys.90.035006}%
  \BibitemOpen
  \bibfield  {author} {\bibinfo {author} {\bibfnamefont {D.}~\bibnamefont {Braun}}, \bibinfo {author} {\bibfnamefont {G.}~\bibnamefont {Adesso}}, \bibinfo {author} {\bibfnamefont {F.}~\bibnamefont {Benatti}}, \bibinfo {author} {\bibfnamefont {R.}~\bibnamefont {Floreanini}}, \bibinfo {author} {\bibfnamefont {U.}~\bibnamefont {Marzolino}}, \bibinfo {author} {\bibfnamefont {M.~W.}\ \bibnamefont {Mitchell}},\ and\ \bibinfo {author} {\bibfnamefont {S.}~\bibnamefont {Pirandola}},\ }\bibfield  {title} {\bibinfo {title} {Quantum-enhanced measurements without entanglement},\ }\href {https://doi.org/10.1103/RevModPhys.90.035006} {\bibfield  {journal} {\bibinfo  {journal} {Rev. Mod. Phys.}\ }\textbf {\bibinfo {volume} {90}},\ \bibinfo {pages} {035006} (\bibinfo {year} {2018})}\BibitemShut {NoStop}%
\bibitem [{\citenamefont {Pirandola}\ \emph {et~al.}(2018)\citenamefont {Pirandola}, \citenamefont {Bardhan}, \citenamefont {Gehring}, \citenamefont {Weedbrook},\ and\ \citenamefont {Lloyd}}]{Pirandola2018}%
  \BibitemOpen
  \bibfield  {author} {\bibinfo {author} {\bibfnamefont {S.}~\bibnamefont {Pirandola}}, \bibinfo {author} {\bibfnamefont {B.~R.}\ \bibnamefont {Bardhan}}, \bibinfo {author} {\bibfnamefont {T.}~\bibnamefont {Gehring}}, \bibinfo {author} {\bibfnamefont {C.}~\bibnamefont {Weedbrook}},\ and\ \bibinfo {author} {\bibfnamefont {S.}~\bibnamefont {Lloyd}},\ }\bibfield  {title} {\bibinfo {title} {Advances in photonic quantum sensing},\ }\href {https://doi.org/10.1038/s41566-018-0301-6} {\bibfield  {journal} {\bibinfo  {journal} {Nat. Photon.}\ }\textbf {\bibinfo {volume} {12}},\ \bibinfo {pages} {724} (\bibinfo {year} {2018})}\BibitemShut {NoStop}%
\bibitem [{\citenamefont {Sahoo}\ \emph {et~al.}(2024)\citenamefont {Sahoo}, \citenamefont {Mishra},\ and\ \citenamefont {Rakshit}}]{PhysRevA.109.L030601}%
  \BibitemOpen
  \bibfield  {author} {\bibinfo {author} {\bibfnamefont {A.}~\bibnamefont {Sahoo}}, \bibinfo {author} {\bibfnamefont {U.}~\bibnamefont {Mishra}},\ and\ \bibinfo {author} {\bibfnamefont {D.}~\bibnamefont {Rakshit}},\ }\bibfield  {title} {\bibinfo {title} {Localization-driven quantum sensing},\ }\href {https://doi.org/10.1103/PhysRevA.109.L030601} {\bibfield  {journal} {\bibinfo  {journal} {Phys. Rev. A}\ }\textbf {\bibinfo {volume} {109}},\ \bibinfo {pages} {L030601} (\bibinfo {year} {2024})}\BibitemShut {NoStop}%
\bibitem [{\citenamefont {Bhattacharyya}\ \emph {et~al.}(2024{\natexlab{a}})\citenamefont {Bhattacharyya}, \citenamefont {Ghoshal},\ and\ \citenamefont {Sen}}]{PhysRevA.109.052626}%
  \BibitemOpen
  \bibfield  {author} {\bibinfo {author} {\bibfnamefont {A.}~\bibnamefont {Bhattacharyya}}, \bibinfo {author} {\bibfnamefont {A.}~\bibnamefont {Ghoshal}},\ and\ \bibinfo {author} {\bibfnamefont {U.}~\bibnamefont {Sen}},\ }\bibfield  {title} {\bibinfo {title} {Restoring metrological quantum advantage of measurement precision in a noisy scenario},\ }\href {https://doi.org/10.1103/PhysRevA.109.052626} {\bibfield  {journal} {\bibinfo  {journal} {Phys. Rev. A}\ }\textbf {\bibinfo {volume} {109}},\ \bibinfo {pages} {052626} (\bibinfo {year} {2024}{\natexlab{a}})}\BibitemShut {NoStop}%
\bibitem [{\citenamefont {Bhattacharyya}\ \emph {et~al.}(2024{\natexlab{b}})\citenamefont {Bhattacharyya}, \citenamefont {Ghoshal},\ and\ \citenamefont {Sen}}]{PhysRevA.110.012620}%
  \BibitemOpen
  \bibfield  {author} {\bibinfo {author} {\bibfnamefont {A.}~\bibnamefont {Bhattacharyya}}, \bibinfo {author} {\bibfnamefont {A.}~\bibnamefont {Ghoshal}},\ and\ \bibinfo {author} {\bibfnamefont {U.}~\bibnamefont {Sen}},\ }\bibfield  {title} {\bibinfo {title} {Enhancing precision of atomic clocks by tuning disorder in accessories},\ }\href {https://doi.org/10.1103/PhysRevA.110.012620} {\bibfield  {journal} {\bibinfo  {journal} {Phys. Rev. A}\ }\textbf {\bibinfo {volume} {110}},\ \bibinfo {pages} {012620} (\bibinfo {year} {2024}{\natexlab{b}})}\BibitemShut {NoStop}%
\bibitem [{\citenamefont {Singh}\ \emph {et~al.}(2024)\citenamefont {Singh}, \citenamefont {Lakkaraju}, \citenamefont {Ghosh},\ and\ \citenamefont {Sen~(De)}}]{singh2024dimensionalgainsensinghigherdimensional}%
  \BibitemOpen
  \bibfield  {author} {\bibinfo {author} {\bibfnamefont {S.}~\bibnamefont {Singh}}, \bibinfo {author} {\bibfnamefont {L.~G.~C.}\ \bibnamefont {Lakkaraju}}, \bibinfo {author} {\bibfnamefont {S.}~\bibnamefont {Ghosh}},\ and\ \bibinfo {author} {\bibfnamefont {A.}~\bibnamefont {Sen~(De)}},\ }\bibfield  {title} {\bibinfo {title} {Dimensional gain in sensing through higher-dimensional quantum spin chain},\ }\href {https://arxiv.org/abs/2401.14853} {\bibfield  {journal} {\bibinfo  {journal} {arXiv:2401.14853}\ } (\bibinfo {year} {2024})}\BibitemShut {NoStop}%
\bibitem [{\citenamefont {Mondal}\ \emph {et~al.}(2024)\citenamefont {Mondal}, \citenamefont {Sahoo}, \citenamefont {Sen},\ and\ \citenamefont {Rakshit}}]{mondal2024multicriticalquantumsensorsdriven}%
  \BibitemOpen
  \bibfield  {author} {\bibinfo {author} {\bibfnamefont {S.}~\bibnamefont {Mondal}}, \bibinfo {author} {\bibfnamefont {A.}~\bibnamefont {Sahoo}}, \bibinfo {author} {\bibfnamefont {U.}~\bibnamefont {Sen}},\ and\ \bibinfo {author} {\bibfnamefont {D.}~\bibnamefont {Rakshit}},\ }\bibfield  {title} {\bibinfo {title} {Multicritical quantum sensors driven by symmetry-breaking},\ }\href {https://arxiv.org/abs/2407.14428} {\bibfield  {journal} {\bibinfo  {journal} {arXiv:2407.14428}\ } (\bibinfo {year} {2024})}\BibitemShut {NoStop}%
\bibitem [{\citenamefont {Sahoo}\ and\ \citenamefont {Rakshit}(2024)}]{sahoo2024enhancedsensingstarkweak}%
  \BibitemOpen
  \bibfield  {author} {\bibinfo {author} {\bibfnamefont {A.}~\bibnamefont {Sahoo}}\ and\ \bibinfo {author} {\bibfnamefont {D.}~\bibnamefont {Rakshit}},\ }\bibfield  {title} {\bibinfo {title} {Enhanced sensing of stark weak field under the influence of aubry-andr\'e-harper criticality},\ }\href {https://arxiv.org/abs/2408.03232} {\bibfield  {journal} {\bibinfo  {journal} {arXiv:2408.03232}\ } (\bibinfo {year} {2024})}\BibitemShut {NoStop}%
\bibitem [{\citenamefont {Mothsara}\ \emph {et~al.}(2025)\citenamefont {Mothsara}, \citenamefont {Lakkaraju}, \citenamefont {Ghosh},\ and\ \citenamefont {Sen~(De)}}]{PhysRevA.111.042628}%
  \BibitemOpen
  \bibfield  {author} {\bibinfo {author} {\bibfnamefont {M.}~\bibnamefont {Mothsara}}, \bibinfo {author} {\bibfnamefont {L.~G.~C.}\ \bibnamefont {Lakkaraju}}, \bibinfo {author} {\bibfnamefont {S.}~\bibnamefont {Ghosh}},\ and\ \bibinfo {author} {\bibfnamefont {A.}~\bibnamefont {Sen~(De)}},\ }\bibfield  {title} {\bibinfo {title} {Quantum-enhanced sensing with variable-range interactions},\ }\href {https://doi.org/10.1103/PhysRevA.111.042628} {\bibfield  {journal} {\bibinfo  {journal} {Phys. Rev. A}\ }\textbf {\bibinfo {volume} {111}},\ \bibinfo {pages} {042628} (\bibinfo {year} {2025})}\BibitemShut {NoStop}%
\bibitem [{\citenamefont {Agrawal}\ \emph {et~al.}(2025)\citenamefont {Agrawal}, \citenamefont {Halder},\ and\ \citenamefont {Sen~(De)}}]{Agrawal2025indefinitetime}%
  \BibitemOpen
  \bibfield  {author} {\bibinfo {author} {\bibfnamefont {G.}~\bibnamefont {Agrawal}}, \bibinfo {author} {\bibfnamefont {P.}~\bibnamefont {Halder}},\ and\ \bibinfo {author} {\bibfnamefont {A.}~\bibnamefont {Sen~(De)}},\ }\bibfield  {title} {\bibinfo {title} {Indefinite {T}ime {D}irected {Q}uantum {M}etrology},\ }\href {https://doi.org/10.22331/q-2025-07-03-1785} {\bibfield  {journal} {\bibinfo  {journal} {{Quantum}}\ }\textbf {\bibinfo {volume} {9}},\ \bibinfo {pages} {1785} (\bibinfo {year} {2025})}\BibitemShut {NoStop}%
\bibitem [{\citenamefont {Bhattacharyya}\ and\ \citenamefont {Sen}(2025)}]{bhattacharyya2025precisionestimatingindependentlocal}%
  \BibitemOpen
  \bibfield  {author} {\bibinfo {author} {\bibfnamefont {A.}~\bibnamefont {Bhattacharyya}}\ and\ \bibinfo {author} {\bibfnamefont {U.}~\bibnamefont {Sen}},\ }\bibfield  {title} {\bibinfo {title} {Precision in estimating independent local fields: attainable bound and indispensability of genuine multiparty entanglement},\ }\href {https://arxiv.org/abs/2407.20142} {\bibfield  {journal} {\bibinfo  {journal} {arXiv:2407.20142}\ } (\bibinfo {year} {2025})}\BibitemShut {NoStop}%
\bibitem [{\citenamefont {Agarwal}\ \emph {et~al.}(2025{\natexlab{a}})\citenamefont {Agarwal}, \citenamefont {Konar}, \citenamefont {Lakkaraju},\ and\ \citenamefont {Sen~(De)}}]{agarwal2025criticalquantummetrologyusing}%
  \BibitemOpen
  \bibfield  {author} {\bibinfo {author} {\bibfnamefont {K.~D.}\ \bibnamefont {Agarwal}}, \bibinfo {author} {\bibfnamefont {T.~K.}\ \bibnamefont {Konar}}, \bibinfo {author} {\bibfnamefont {L.~G.~C.}\ \bibnamefont {Lakkaraju}},\ and\ \bibinfo {author} {\bibfnamefont {A.}~\bibnamefont {Sen~(De)}},\ }\bibfield  {title} {\bibinfo {title} {Critical quantum metrology using non-hermitian spin model with rt-symmetry},\ }\href {https://arxiv.org/abs/2503.24331} {\bibfield  {journal} {\bibinfo  {journal} {arXiv:2503.24331}\ } (\bibinfo {year} {2025}{\natexlab{a}})}\BibitemShut {NoStop}%
\bibitem [{\citenamefont {Bhattacharyya}\ \emph {et~al.}(2025{\natexlab{a}})\citenamefont {Bhattacharyya}, \citenamefont {Saha},\ and\ \citenamefont {Sen}}]{bhattacharyya2025quantumsensingevenversus}%
  \BibitemOpen
  \bibfield  {author} {\bibinfo {author} {\bibfnamefont {A.}~\bibnamefont {Bhattacharyya}}, \bibinfo {author} {\bibfnamefont {D.}~\bibnamefont {Saha}},\ and\ \bibinfo {author} {\bibfnamefont {U.}~\bibnamefont {Sen}},\ }\bibfield  {title} {\bibinfo {title} {Quantum sensing of even- versus odd-body interactions},\ }\href {https://arxiv.org/abs/2401.06729} {\bibfield  {journal} {\bibinfo  {journal} {arXiv: 2401.06729}\ } (\bibinfo {year} {2025}{\natexlab{a}})}\BibitemShut {NoStop}%
\bibitem [{\citenamefont {Saha}\ and\ \citenamefont {Sen}(2025)}]{saha2025entanglementconstrainedquantummetrologyrapid}%
  \BibitemOpen
  \bibfield  {author} {\bibinfo {author} {\bibfnamefont {D.}~\bibnamefont {Saha}}\ and\ \bibinfo {author} {\bibfnamefont {U.}~\bibnamefont {Sen}},\ }\bibfield  {title} {\bibinfo {title} {Entanglement-constrained quantum metrology: Rapid low-entanglement gains, tapered high-level growth},\ }\href {https://arxiv.org/abs/2507.03512} {\bibfield  {journal} {\bibinfo  {journal} {arXiv:2507.03512}\ } (\bibinfo {year} {2025})}\BibitemShut {NoStop}%
\bibitem [{\citenamefont {Pal}\ \emph {et~al.}(2025)\citenamefont {Pal}, \citenamefont {Ghosh}, \citenamefont {Ghoshal},\ and\ \citenamefont {Sen}}]{pal2025rolephaseoptimalprobe}%
  \BibitemOpen
  \bibfield  {author} {\bibinfo {author} {\bibfnamefont {R.}~\bibnamefont {Pal}}, \bibinfo {author} {\bibfnamefont {P.}~\bibnamefont {Ghosh}}, \bibinfo {author} {\bibfnamefont {A.}~\bibnamefont {Ghoshal}},\ and\ \bibinfo {author} {\bibfnamefont {U.}~\bibnamefont {Sen}},\ }\bibfield  {title} {\bibinfo {title} {Role of phase of optimal probe in noncommutativity vs coherence in quantum multiparameter estimation},\ }\href {https://arxiv.org/abs/2507.04824} {\bibfield  {journal} {\bibinfo  {journal} {arXiv:2507.04824}\ } (\bibinfo {year} {2025})}\BibitemShut {NoStop}%
\bibitem [{\citenamefont {Agarwal}\ \emph {et~al.}(2025{\natexlab{b}})\citenamefont {Agarwal}, \citenamefont {Mondal}, \citenamefont {Sahoo}, \citenamefont {Rakshit}, \citenamefont {Sen~(De)},\ and\ \citenamefont {Sen}}]{agarwal2025quantumsensingultracoldsimulators}%
  \BibitemOpen
  \bibfield  {author} {\bibinfo {author} {\bibfnamefont {K.~D.}\ \bibnamefont {Agarwal}}, \bibinfo {author} {\bibfnamefont {S.}~\bibnamefont {Mondal}}, \bibinfo {author} {\bibfnamefont {A.}~\bibnamefont {Sahoo}}, \bibinfo {author} {\bibfnamefont {D.}~\bibnamefont {Rakshit}}, \bibinfo {author} {\bibfnamefont {A.}~\bibnamefont {Sen~(De)}},\ and\ \bibinfo {author} {\bibfnamefont {U.}~\bibnamefont {Sen}},\ }\bibfield  {title} {\bibinfo {title} {Quantum sensing with ultracold simulators in lattice and ensemble systems: a review},\ }\href {https://arxiv.org/abs/2507.06348} {\bibfield  {journal} {\bibinfo  {journal} {arXiv:2507.06348}\ } (\bibinfo {year} {2025}{\natexlab{b}})}\BibitemShut {NoStop}%
\bibitem [{\citenamefont {Mondal}\ \emph {et~al.}(2025)\citenamefont {Mondal}, \citenamefont {Ghosh},\ and\ \citenamefont {Sen}}]{mondal2025optimalquantumprecisionnoise}%
  \BibitemOpen
  \bibfield  {author} {\bibinfo {author} {\bibfnamefont {S.}~\bibnamefont {Mondal}}, \bibinfo {author} {\bibfnamefont {P.}~\bibnamefont {Ghosh}},\ and\ \bibinfo {author} {\bibfnamefont {U.}~\bibnamefont {Sen}},\ }\bibfield  {title} {\bibinfo {title} {Optimal quantum precision in noise estimation: Is entanglement necessary?},\ }\href {https://arxiv.org/abs/2507.22413} {\bibfield  {journal} {\bibinfo  {journal} {arXiv:2507.22413}\ } (\bibinfo {year} {2025})}\BibitemShut {NoStop}%
\bibitem [{\citenamefont {Chaki}\ \emph {et~al.}(2025)\citenamefont {Chaki}, \citenamefont {Saha}, \citenamefont {Sen},\ and\ \citenamefont {Sen}}]{chaki2025nonpositivemeasurementsarentbeneficial}%
  \BibitemOpen
  \bibfield  {author} {\bibinfo {author} {\bibfnamefont {P.}~\bibnamefont {Chaki}}, \bibinfo {author} {\bibfnamefont {D.}~\bibnamefont {Saha}}, \bibinfo {author} {\bibfnamefont {K.}~\bibnamefont {Sen}},\ and\ \bibinfo {author} {\bibfnamefont {U.}~\bibnamefont {Sen}},\ }\bibfield  {title} {\bibinfo {title} {Non-positive measurements aren't beneficial in quantum metrology for unitary encoding, but can be for open schemes},\ }\href {https://arxiv.org/abs/2509.24585} {\bibfield  {journal} {\bibinfo  {journal} {arXiv:2509.24585}\ } (\bibinfo {year} {2025})}\BibitemShut {NoStop}%
\bibitem [{\citenamefont {Nagata}\ \emph {et~al.}(2007)\citenamefont {Nagata}, \citenamefont {Okamoto}, \citenamefont {O'Brien}, \citenamefont {Sasaki},\ and\ \citenamefont {Takeuchi}}]{nagata}%
  \BibitemOpen
  \bibfield  {author} {\bibinfo {author} {\bibfnamefont {T.}~\bibnamefont {Nagata}}, \bibinfo {author} {\bibfnamefont {R.}~\bibnamefont {Okamoto}}, \bibinfo {author} {\bibfnamefont {J.~L.}\ \bibnamefont {O'Brien}}, \bibinfo {author} {\bibfnamefont {K.}~\bibnamefont {Sasaki}},\ and\ \bibinfo {author} {\bibfnamefont {S.}~\bibnamefont {Takeuchi}},\ }\bibfield  {title} {\bibinfo {title} {Beating the standard quantum limit with four-entangled photons},\ }\href {https://www.science.org/doi/abs/10.1126/science.1138007} {\bibfield  {journal} {\bibinfo  {journal} {Science}\ }\textbf {\bibinfo {volume} {316}},\ \bibinfo {pages} {726} (\bibinfo {year} {2007})}\BibitemShut {NoStop}%
\bibitem [{\citenamefont {Gross}(2012)}]{Gross_2012}%
  \BibitemOpen
  \bibfield  {author} {\bibinfo {author} {\bibfnamefont {C.}~\bibnamefont {Gross}},\ }\bibfield  {title} {\bibinfo {title} {Spin squeezing, entanglement and quantum metrology with bose–einstein condensates},\ }\href {https://doi.org/10.1088/0953-4075/45/10/103001} {\bibfield  {journal} {\bibinfo  {journal} {J. Phys. B: At. Mol. Opt. Phys.}\ }\textbf {\bibinfo {volume} {45}},\ \bibinfo {pages} {103001} (\bibinfo {year} {2012})}\BibitemShut {NoStop}%
\bibitem [{\citenamefont {Israel}\ \emph {et~al.}(2014)\citenamefont {Israel}, \citenamefont {Rosen},\ and\ \citenamefont {Silberberg}}]{Israel2014}%
  \BibitemOpen
  \bibfield  {author} {\bibinfo {author} {\bibfnamefont {Y.}~\bibnamefont {Israel}}, \bibinfo {author} {\bibfnamefont {S.}~\bibnamefont {Rosen}},\ and\ \bibinfo {author} {\bibfnamefont {Y.}~\bibnamefont {Silberberg}},\ }\bibfield  {title} {\bibinfo {title} {Supersensitive polarization microscopy using noon states of light},\ }\href {https://doi.org/10.1103/PhysRevLett.112.103604} {\bibfield  {journal} {\bibinfo  {journal} {Phys. Rev. Lett.}\ }\textbf {\bibinfo {volume} {112}},\ \bibinfo {pages} {103604} (\bibinfo {year} {2014})}\BibitemShut {NoStop}%
\bibitem [{\citenamefont {Linden}\ \emph {et~al.}(2010)\citenamefont {Linden}, \citenamefont {Popescu},\ and\ \citenamefont {Skrzypczyk}}]{PhysRevLett.105.130401}%
  \BibitemOpen
  \bibfield  {author} {\bibinfo {author} {\bibfnamefont {N.}~\bibnamefont {Linden}}, \bibinfo {author} {\bibfnamefont {S.}~\bibnamefont {Popescu}},\ and\ \bibinfo {author} {\bibfnamefont {P.}~\bibnamefont {Skrzypczyk}},\ }\bibfield  {title} {\bibinfo {title} {How small can thermal machines be? the smallest possible refrigerator},\ }\href {https://doi.org/10.1103/PhysRevLett.105.130401} {\bibfield  {journal} {\bibinfo  {journal} {Phys. Rev. Lett.}\ }\textbf {\bibinfo {volume} {105}},\ \bibinfo {pages} {130401} (\bibinfo {year} {2010})}\BibitemShut {NoStop}%
\bibitem [{\citenamefont {Skrzypczyk}\ \emph {et~al.}(2011)\citenamefont {Skrzypczyk}, \citenamefont {Brunner}, \citenamefont {Linden},\ and\ \citenamefont {Popescu}}]{Skrzypczyk_2011}%
  \BibitemOpen
  \bibfield  {author} {\bibinfo {author} {\bibfnamefont {P.}~\bibnamefont {Skrzypczyk}}, \bibinfo {author} {\bibfnamefont {N.}~\bibnamefont {Brunner}}, \bibinfo {author} {\bibfnamefont {N.}~\bibnamefont {Linden}},\ and\ \bibinfo {author} {\bibfnamefont {S.}~\bibnamefont {Popescu}},\ }\bibfield  {title} {\bibinfo {title} {The smallest refrigerators can reach maximal efficiency},\ }\href {https://doi.org/10.1088/1751-8113/44/49/492002} {\bibfield  {journal} {\bibinfo  {journal} {J. Phys. A: Math. Theor.}\ }\textbf {\bibinfo {volume} {44}},\ \bibinfo {pages} {492002} (\bibinfo {year} {2011})}\BibitemShut {NoStop}%
\bibitem [{\citenamefont {Levy}\ and\ \citenamefont {Kosloff}(2012)}]{PhysRevLett.108.070604}%
  \BibitemOpen
  \bibfield  {author} {\bibinfo {author} {\bibfnamefont {A.}~\bibnamefont {Levy}}\ and\ \bibinfo {author} {\bibfnamefont {R.}~\bibnamefont {Kosloff}},\ }\bibfield  {title} {\bibinfo {title} {Quantum absorption refrigerator},\ }\href {https://doi.org/10.1103/PhysRevLett.108.070604} {\bibfield  {journal} {\bibinfo  {journal} {Phys. Rev. Lett.}\ }\textbf {\bibinfo {volume} {108}},\ \bibinfo {pages} {070604} (\bibinfo {year} {2012})}\BibitemShut {NoStop}%
\bibitem [{\citenamefont {Correa}\ \emph {et~al.}(2014)\citenamefont {Correa}, \citenamefont {Palao}, \citenamefont {Alonso},\ and\ \citenamefont {Adesso}}]{Correa2014}%
  \BibitemOpen
  \bibfield  {author} {\bibinfo {author} {\bibfnamefont {L.~A.}\ \bibnamefont {Correa}}, \bibinfo {author} {\bibfnamefont {J.~P.}\ \bibnamefont {Palao}}, \bibinfo {author} {\bibfnamefont {D.}~\bibnamefont {Alonso}},\ and\ \bibinfo {author} {\bibfnamefont {G.}~\bibnamefont {Adesso}},\ }\bibfield  {title} {\bibinfo {title} {Quantum-enhanced absorption refrigerators},\ }\href {https://doi.org/10.1038/srep03949} {\bibfield  {journal} {\bibinfo  {journal} {Sci. Rep.}\ }\textbf {\bibinfo {volume} {4}},\ \bibinfo {pages} {3949} (\bibinfo {year} {2014})}\BibitemShut {NoStop}%
\bibitem [{\citenamefont {Brunner}\ \emph {et~al.}(2014)\citenamefont {Brunner}, \citenamefont {Huber}, \citenamefont {Linden}, \citenamefont {Popescu}, \citenamefont {Silva},\ and\ \citenamefont {Skrzypczyk}}]{PhysRevE.89.032115}%
  \BibitemOpen
  \bibfield  {author} {\bibinfo {author} {\bibfnamefont {N.}~\bibnamefont {Brunner}}, \bibinfo {author} {\bibfnamefont {M.}~\bibnamefont {Huber}}, \bibinfo {author} {\bibfnamefont {N.}~\bibnamefont {Linden}}, \bibinfo {author} {\bibfnamefont {S.}~\bibnamefont {Popescu}}, \bibinfo {author} {\bibfnamefont {R.}~\bibnamefont {Silva}},\ and\ \bibinfo {author} {\bibfnamefont {P.}~\bibnamefont {Skrzypczyk}},\ }\bibfield  {title} {\bibinfo {title} {Entanglement enhances cooling in microscopic quantum refrigerators},\ }\href {https://doi.org/10.1103/PhysRevE.89.032115} {\bibfield  {journal} {\bibinfo  {journal} {Phys. Rev. E}\ }\textbf {\bibinfo {volume} {89}},\ \bibinfo {pages} {032115} (\bibinfo {year} {2014})}\BibitemShut {NoStop}%
\bibitem [{\citenamefont {Wang}\ \emph {et~al.}(2015)\citenamefont {Wang}, \citenamefont {Lai}, \citenamefont {Ye}, \citenamefont {He}, \citenamefont {Ma},\ and\ \citenamefont {Liao}}]{PhysRevE.91.050102}%
  \BibitemOpen
  \bibfield  {author} {\bibinfo {author} {\bibfnamefont {J.}~\bibnamefont {Wang}}, \bibinfo {author} {\bibfnamefont {Y.}~\bibnamefont {Lai}}, \bibinfo {author} {\bibfnamefont {Z.}~\bibnamefont {Ye}}, \bibinfo {author} {\bibfnamefont {J.}~\bibnamefont {He}}, \bibinfo {author} {\bibfnamefont {Y.}~\bibnamefont {Ma}},\ and\ \bibinfo {author} {\bibfnamefont {Q.}~\bibnamefont {Liao}},\ }\bibfield  {title} {\bibinfo {title} {Four-level refrigerator driven by photons},\ }\href {https://doi.org/10.1103/PhysRevE.91.050102} {\bibfield  {journal} {\bibinfo  {journal} {Phys. Rev. E}\ }\textbf {\bibinfo {volume} {91}},\ \bibinfo {pages} {050102} (\bibinfo {year} {2015})}\BibitemShut {NoStop}%
\bibitem [{\citenamefont {Mitchison}\ \emph {et~al.}(2015)\citenamefont {Mitchison}, \citenamefont {Woods}, \citenamefont {Prior},\ and\ \citenamefont {Huber}}]{Mitchison_2015}%
  \BibitemOpen
  \bibfield  {author} {\bibinfo {author} {\bibfnamefont {M.~T.}\ \bibnamefont {Mitchison}}, \bibinfo {author} {\bibfnamefont {M.~P.}\ \bibnamefont {Woods}}, \bibinfo {author} {\bibfnamefont {J.}~\bibnamefont {Prior}},\ and\ \bibinfo {author} {\bibfnamefont {M.}~\bibnamefont {Huber}},\ }\bibfield  {title} {\bibinfo {title} {Coherence-assisted single-shot cooling by quantum absorption refrigerators},\ }\href {https://doi.org/10.1088/1367-2630/17/11/115013} {\bibfield  {journal} {\bibinfo  {journal} {New J. Phys.}\ }\textbf {\bibinfo {volume} {17}},\ \bibinfo {pages} {115013} (\bibinfo {year} {2015})}\BibitemShut {NoStop}%
\bibitem [{\citenamefont {Brask}\ and\ \citenamefont {Brunner}(2015)}]{PhysRevE.92.062101}%
  \BibitemOpen
  \bibfield  {author} {\bibinfo {author} {\bibfnamefont {J.~B.}\ \bibnamefont {Brask}}\ and\ \bibinfo {author} {\bibfnamefont {N.}~\bibnamefont {Brunner}},\ }\bibfield  {title} {\bibinfo {title} {Small quantum absorption refrigerator in the transient regime: Time scales, enhanced cooling, and entanglement},\ }\href {https://doi.org/10.1103/PhysRevE.92.062101} {\bibfield  {journal} {\bibinfo  {journal} {Phys. Rev. E}\ }\textbf {\bibinfo {volume} {92}},\ \bibinfo {pages} {062101} (\bibinfo {year} {2015})}\BibitemShut {NoStop}%
\bibitem [{\citenamefont {Nimmrichter}\ \emph {et~al.}(2017)\citenamefont {Nimmrichter}, \citenamefont {Dai}, \citenamefont {Roulet},\ and\ \citenamefont {Scarani}}]{Nimmrichter2017quantumclassical}%
  \BibitemOpen
  \bibfield  {author} {\bibinfo {author} {\bibfnamefont {S.}~\bibnamefont {Nimmrichter}}, \bibinfo {author} {\bibfnamefont {J.}~\bibnamefont {Dai}}, \bibinfo {author} {\bibfnamefont {A.}~\bibnamefont {Roulet}},\ and\ \bibinfo {author} {\bibfnamefont {V.}~\bibnamefont {Scarani}},\ }\bibfield  {title} {\bibinfo {title} {Quantum and classical dynamics of a three-mode absorption refrigerator},\ }\href {https://doi.org/10.22331/q-2017-12-11-37} {\bibfield  {journal} {\bibinfo  {journal} {{Quantum}}\ }\textbf {\bibinfo {volume} {1}},\ \bibinfo {pages} {37} (\bibinfo {year} {2017})}\BibitemShut {NoStop}%
\bibitem [{\citenamefont {Mu}\ \emph {et~al.}(2017)\citenamefont {Mu}, \citenamefont {Agarwalla}, \citenamefont {Schaller},\ and\ \citenamefont {Segal}}]{Mu_2017}%
  \BibitemOpen
  \bibfield  {author} {\bibinfo {author} {\bibfnamefont {A.}~\bibnamefont {Mu}}, \bibinfo {author} {\bibfnamefont {B.~K.}\ \bibnamefont {Agarwalla}}, \bibinfo {author} {\bibfnamefont {G.}~\bibnamefont {Schaller}},\ and\ \bibinfo {author} {\bibfnamefont {D.}~\bibnamefont {Segal}},\ }\bibfield  {title} {\bibinfo {title} {Qubit absorption refrigerator at strong coupling},\ }\href {https://doi.org/10.1088/1367-2630/aa9b75} {\bibfield  {journal} {\bibinfo  {journal} {New J. Phys.}\ }\textbf {\bibinfo {volume} {19}},\ \bibinfo {pages} {123034} (\bibinfo {year} {2017})}\BibitemShut {NoStop}%
\bibitem [{\citenamefont {Mukhopadhyay}\ \emph {et~al.}(2018)\citenamefont {Mukhopadhyay}, \citenamefont {Misra}, \citenamefont {Bhattacharya},\ and\ \citenamefont {Pati}}]{PhysRevE.97.062116}%
  \BibitemOpen
  \bibfield  {author} {\bibinfo {author} {\bibfnamefont {C.}~\bibnamefont {Mukhopadhyay}}, \bibinfo {author} {\bibfnamefont {A.}~\bibnamefont {Misra}}, \bibinfo {author} {\bibfnamefont {S.}~\bibnamefont {Bhattacharya}},\ and\ \bibinfo {author} {\bibfnamefont {A.~K.}\ \bibnamefont {Pati}},\ }\bibfield  {title} {\bibinfo {title} {Quantum speed limit constraints on a nanoscale autonomous refrigerator},\ }\href {https://doi.org/10.1103/PhysRevE.97.062116} {\bibfield  {journal} {\bibinfo  {journal} {Phys. Rev. E}\ }\textbf {\bibinfo {volume} {97}},\ \bibinfo {pages} {062116} (\bibinfo {year} {2018})}\BibitemShut {NoStop}%
\bibitem [{\citenamefont {Das}\ \emph {et~al.}(2019)\citenamefont {Das}, \citenamefont {Misra}, \citenamefont {Pal}, \citenamefont {Sen~(De)},\ and\ \citenamefont {Sen}}]{Das_2019}%
  \BibitemOpen
  \bibfield  {author} {\bibinfo {author} {\bibfnamefont {S.}~\bibnamefont {Das}}, \bibinfo {author} {\bibfnamefont {A.}~\bibnamefont {Misra}}, \bibinfo {author} {\bibfnamefont {A.~K.}\ \bibnamefont {Pal}}, \bibinfo {author} {\bibfnamefont {A.}~\bibnamefont {Sen~(De)}},\ and\ \bibinfo {author} {\bibfnamefont {U.}~\bibnamefont {Sen}},\ }\bibfield  {title} {\bibinfo {title} {Necessarily transient quantum refrigerator},\ }\href {https://doi.org/10.1209/0295-5075/125/20007} {\bibfield  {journal} {\bibinfo  {journal} {Europhys. Lett.}\ }\textbf {\bibinfo {volume} {125}},\ \bibinfo {pages} {20007} (\bibinfo {year} {2019})}\BibitemShut {NoStop}%
\bibitem [{\citenamefont {Mitchison}(2019)}]{Mitchison03042019}%
  \BibitemOpen
  \bibfield  {author} {\bibinfo {author} {\bibfnamefont {M.~T.}\ \bibnamefont {Mitchison}},\ }\bibfield  {title} {\bibinfo {title} {Quantum thermal absorption machines: refrigerators, engines and clocks},\ }\href {https://doi.org/10.1080/00107514.2019.1631555} {\bibfield  {journal} {\bibinfo  {journal} {Contemp. Phys.}\ }\textbf {\bibinfo {volume} {60}},\ \bibinfo {pages} {164} (\bibinfo {year} {2019})}\BibitemShut {NoStop}%
\bibitem [{\citenamefont {Hewgill}\ \emph {et~al.}(2020)\citenamefont {Hewgill}, \citenamefont {Gonz\'alez}, \citenamefont {Palao}, \citenamefont {Alonso}, \citenamefont {Ferraro},\ and\ \citenamefont {De~Chiara}}]{PhysRevE.101.012109}%
  \BibitemOpen
  \bibfield  {author} {\bibinfo {author} {\bibfnamefont {A.}~\bibnamefont {Hewgill}}, \bibinfo {author} {\bibfnamefont {J.~O.}\ \bibnamefont {Gonz\'alez}}, \bibinfo {author} {\bibfnamefont {J.~P.}\ \bibnamefont {Palao}}, \bibinfo {author} {\bibfnamefont {D.}~\bibnamefont {Alonso}}, \bibinfo {author} {\bibfnamefont {A.}~\bibnamefont {Ferraro}},\ and\ \bibinfo {author} {\bibfnamefont {G.}~\bibnamefont {De~Chiara}},\ }\bibfield  {title} {\bibinfo {title} {Three-qubit refrigerator with two-body interactions},\ }\href {https://doi.org/10.1103/PhysRevE.101.012109} {\bibfield  {journal} {\bibinfo  {journal} {Phys. Rev. E}\ }\textbf {\bibinfo {volume} {101}},\ \bibinfo {pages} {012109} (\bibinfo {year} {2020})}\BibitemShut {NoStop}%
\bibitem [{\citenamefont {Ghoshal}\ \emph {et~al.}(2021)\citenamefont {Ghoshal}, \citenamefont {Das}, \citenamefont {Pal}, \citenamefont {Sen~(De)},\ and\ \citenamefont {Sen}}]{PhysRevA.104.042208}%
  \BibitemOpen
  \bibfield  {author} {\bibinfo {author} {\bibfnamefont {A.}~\bibnamefont {Ghoshal}}, \bibinfo {author} {\bibfnamefont {S.}~\bibnamefont {Das}}, \bibinfo {author} {\bibfnamefont {A.~K.}\ \bibnamefont {Pal}}, \bibinfo {author} {\bibfnamefont {A.}~\bibnamefont {Sen~(De)}},\ and\ \bibinfo {author} {\bibfnamefont {U.}~\bibnamefont {Sen}},\ }\bibfield  {title} {\bibinfo {title} {Three qubits in less than three baths: Beyond two-body system-bath interactions in quantum refrigerators},\ }\href {https://doi.org/10.1103/PhysRevA.104.042208} {\bibfield  {journal} {\bibinfo  {journal} {Phys. Rev. A}\ }\textbf {\bibinfo {volume} {104}},\ \bibinfo {pages} {042208} (\bibinfo {year} {2021})}\BibitemShut {NoStop}%
\bibitem [{\citenamefont {Konar}\ \emph {et~al.}(2022{\natexlab{a}})\citenamefont {Konar}, \citenamefont {Ghosh}, \citenamefont {Pal},\ and\ \citenamefont {Sen~(De)}}]{PhysRevA.105.022214}%
  \BibitemOpen
  \bibfield  {author} {\bibinfo {author} {\bibfnamefont {T.~K.}\ \bibnamefont {Konar}}, \bibinfo {author} {\bibfnamefont {S.}~\bibnamefont {Ghosh}}, \bibinfo {author} {\bibfnamefont {A.~K.}\ \bibnamefont {Pal}},\ and\ \bibinfo {author} {\bibfnamefont {A.}~\bibnamefont {Sen~(De)}},\ }\bibfield  {title} {\bibinfo {title} {Designing robust quantum refrigerators in disordered spin models},\ }\href {https://doi.org/10.1103/PhysRevA.105.022214} {\bibfield  {journal} {\bibinfo  {journal} {Phys. Rev. A}\ }\textbf {\bibinfo {volume} {105}},\ \bibinfo {pages} {022214} (\bibinfo {year} {2022}{\natexlab{a}})}\BibitemShut {NoStop}%
\bibitem [{\citenamefont {Konar}\ \emph {et~al.}(2022{\natexlab{b}})\citenamefont {Konar}, \citenamefont {Ghosh},\ and\ \citenamefont {Sen~(De)}}]{PhysRevA.106.022616}%
  \BibitemOpen
  \bibfield  {author} {\bibinfo {author} {\bibfnamefont {T.~K.}\ \bibnamefont {Konar}}, \bibinfo {author} {\bibfnamefont {S.}~\bibnamefont {Ghosh}},\ and\ \bibinfo {author} {\bibfnamefont {A.}~\bibnamefont {Sen~(De)}},\ }\bibfield  {title} {\bibinfo {title} {Refrigeration via purification through repeated measurements},\ }\href {https://doi.org/10.1103/PhysRevA.106.022616} {\bibfield  {journal} {\bibinfo  {journal} {Phys. Rev. A}\ }\textbf {\bibinfo {volume} {106}},\ \bibinfo {pages} {022616} (\bibinfo {year} {2022}{\natexlab{b}})}\BibitemShut {NoStop}%
\bibitem [{\citenamefont {Konar}\ \emph {et~al.}(2023)\citenamefont {Konar}, \citenamefont {Ghosh}, \citenamefont {Pal},\ and\ \citenamefont {Sen~(De)}}]{PhysRevA.107.032602}%
  \BibitemOpen
  \bibfield  {author} {\bibinfo {author} {\bibfnamefont {T.~K.}\ \bibnamefont {Konar}}, \bibinfo {author} {\bibfnamefont {S.}~\bibnamefont {Ghosh}}, \bibinfo {author} {\bibfnamefont {A.~K.}\ \bibnamefont {Pal}},\ and\ \bibinfo {author} {\bibfnamefont {A.}~\bibnamefont {Sen~(De)}},\ }\bibfield  {title} {\bibinfo {title} {Designing refrigerators in higher dimensions using quantum spin models},\ }\href {https://doi.org/10.1103/PhysRevA.107.032602} {\bibfield  {journal} {\bibinfo  {journal} {Phys. Rev. A}\ }\textbf {\bibinfo {volume} {107}},\ \bibinfo {pages} {032602} (\bibinfo {year} {2023})}\BibitemShut {NoStop}%
\bibitem [{\citenamefont {Ray}\ \emph {et~al.}(2023)\citenamefont {Ray}, \citenamefont {Mondal}, \citenamefont {Bhattacharyya}, \citenamefont {Ghoshal}, \citenamefont {Rakshit},\ and\ \citenamefont {Sen}}]{ray2023kerrtypenonlinearbathsenhance}%
  \BibitemOpen
  \bibfield  {author} {\bibinfo {author} {\bibfnamefont {T.}~\bibnamefont {Ray}}, \bibinfo {author} {\bibfnamefont {S.}~\bibnamefont {Mondal}}, \bibinfo {author} {\bibfnamefont {A.}~\bibnamefont {Bhattacharyya}}, \bibinfo {author} {\bibfnamefont {A.}~\bibnamefont {Ghoshal}}, \bibinfo {author} {\bibfnamefont {D.}~\bibnamefont {Rakshit}},\ and\ \bibinfo {author} {\bibfnamefont {U.}~\bibnamefont {Sen}},\ }\bibfield  {title} {\bibinfo {title} {Kerr-type nonlinear baths enhance cooling in quantum refrigerators},\ }\href {https://arxiv.org/abs/2311.10499} {\bibfield  {journal} {\bibinfo  {journal} {arXiv:2311.10499}\ } (\bibinfo {year} {2023})}\BibitemShut {NoStop}%
\bibitem [{\citenamefont {Ghosh}\ \emph {et~al.}(2024)\citenamefont {Ghosh}, \citenamefont {Konar},\ and\ \citenamefont {Sen~(De)}}]{ghosh2024measurementbasedquditquantumrefrigerator}%
  \BibitemOpen
  \bibfield  {author} {\bibinfo {author} {\bibfnamefont {D.}~\bibnamefont {Ghosh}}, \bibinfo {author} {\bibfnamefont {T.~K.}\ \bibnamefont {Konar}},\ and\ \bibinfo {author} {\bibfnamefont {A.}~\bibnamefont {Sen~(De)}},\ }\bibfield  {title} {\bibinfo {title} {Measurement-based qudit quantum refrigerator with subspace cooling},\ }\href {https://arxiv.org/abs/2409.08375} {\bibfield  {journal} {\bibinfo  {journal} {arXiv:2409.08375}\ } (\bibinfo {year} {2024})}\BibitemShut {NoStop}%
\bibitem [{\citenamefont {Bhattacharyya}\ \emph {et~al.}(2025{\natexlab{b}})\citenamefont {Bhattacharyya}, \citenamefont {Ghoshal},\ and\ \citenamefont {Sen}}]{PhysRevA.111.012209}%
  \BibitemOpen
  \bibfield  {author} {\bibinfo {author} {\bibfnamefont {A.}~\bibnamefont {Bhattacharyya}}, \bibinfo {author} {\bibfnamefont {A.}~\bibnamefont {Ghoshal}},\ and\ \bibinfo {author} {\bibfnamefont {U.}~\bibnamefont {Sen}},\ }\bibfield  {title} {\bibinfo {title} {Transient effects in quantum refrigerators with finite environments},\ }\href {https://doi.org/10.1103/PhysRevA.111.012209} {\bibfield  {journal} {\bibinfo  {journal} {Phys. Rev. A}\ }\textbf {\bibinfo {volume} {111}},\ \bibinfo {pages} {012209} (\bibinfo {year} {2025}{\natexlab{b}})}\BibitemShut {NoStop}%
\bibitem [{\citenamefont {Mondkar}\ \emph {et~al.}(2025)\citenamefont {Mondkar}, \citenamefont {Bhattacharyya},\ and\ \citenamefont {Sen}}]{mondkar2025quantumrefrigeratorembeddedspinstar}%
  \BibitemOpen
  \bibfield  {author} {\bibinfo {author} {\bibfnamefont {S.}~\bibnamefont {Mondkar}}, \bibinfo {author} {\bibfnamefont {A.}~\bibnamefont {Bhattacharyya}},\ and\ \bibinfo {author} {\bibfnamefont {U.}~\bibnamefont {Sen}},\ }\bibfield  {title} {\bibinfo {title} {Quantum refrigerator embedded in spin-star environments: Scalings of temperature and refrigeration time},\ }\href {https://arxiv.org/abs/2505.04374} {\bibfield  {journal} {\bibinfo  {journal} {arXiv:2505.04374}\ } (\bibinfo {year} {2025})}\BibitemShut {NoStop}%
\bibitem [{\citenamefont {Mondal}\ and\ \citenamefont {Sen}(2025)}]{mondal2025mpembaeffectselfcontainedquantum}%
  \BibitemOpen
  \bibfield  {author} {\bibinfo {author} {\bibfnamefont {S.}~\bibnamefont {Mondal}}\ and\ \bibinfo {author} {\bibfnamefont {U.}~\bibnamefont {Sen}},\ }\bibfield  {title} {\bibinfo {title} {Mpemba effect in self-contained quantum refrigerators: accelerated cooling},\ }\href {https://arxiv.org/abs/2507.15811} {\bibfield  {journal} {\bibinfo  {journal} {arXiv:2507.15811}\ } (\bibinfo {year} {2025})}\BibitemShut {NoStop}%
\bibitem [{\citenamefont {Jaynes}(1957)}]{PhysRev.106.620}%
  \BibitemOpen
  \bibfield  {author} {\bibinfo {author} {\bibfnamefont {E.~T.}\ \bibnamefont {Jaynes}},\ }\bibfield  {title} {\bibinfo {title} {Information theory and statistical mechanics},\ }\href {https://doi.org/10.1103/PhysRev.106.620} {\bibfield  {journal} {\bibinfo  {journal} {Phys. Rev.}\ }\textbf {\bibinfo {volume} {106}},\ \bibinfo {pages} {620} (\bibinfo {year} {1957})}\BibitemShut {NoStop}%
\bibitem [{\citenamefont {Tagliani}(1998)}]{TAGLIANI1998157}%
  \BibitemOpen
  \bibfield  {author} {\bibinfo {author} {\bibfnamefont {A.}~\bibnamefont {Tagliani}},\ }\bibfield  {title} {\bibinfo {title} {Inverse two-sided laplace transform for probability density functions},\ }\href {https://doi.org/https://doi.org/10.1016/S0377-0427(98)00013-2} {\bibfield  {journal} {\bibinfo  {journal} {J. Comput. Appl. Math.}\ }\textbf {\bibinfo {volume} {90}},\ \bibinfo {pages} {157} (\bibinfo {year} {1998})}\BibitemShut {NoStop}%
\bibitem [{\citenamefont {Bretthorst}(2013)}]{10.1063/1.4819977}%
  \BibitemOpen
  \bibfield  {author} {\bibinfo {author} {\bibfnamefont {G.~L.}\ \bibnamefont {Bretthorst}},\ }\bibfield  {title} {\bibinfo {title} {The maximum entropy method of moments and bayesian probability theory},\ }\href {https://doi.org/10.1063/1.4819977} {\bibfield  {journal} {\bibinfo  {journal} {AIP Conf. Proc.}\ }\textbf {\bibinfo {volume} {1553}},\ \bibinfo {pages} {3} (\bibinfo {year} {2013})}\BibitemShut {NoStop}%
\bibitem [{\citenamefont {Correa}\ \emph {et~al.}(2015)\citenamefont {Correa}, \citenamefont {Mehboudi}, \citenamefont {Adesso},\ and\ \citenamefont {Sanpera}}]{Correa2015-thermometry}%
  \BibitemOpen
  \bibfield  {author} {\bibinfo {author} {\bibfnamefont {L.~A.}\ \bibnamefont {Correa}}, \bibinfo {author} {\bibfnamefont {M.}~\bibnamefont {Mehboudi}}, \bibinfo {author} {\bibfnamefont {G.}~\bibnamefont {Adesso}},\ and\ \bibinfo {author} {\bibfnamefont {A.}~\bibnamefont {Sanpera}},\ }\bibfield  {title} {\bibinfo {title} {Individual quantum probes for optimal thermometry},\ }\href {https://doi.org/10.1103/PhysRevLett.114.220405} {\bibfield  {journal} {\bibinfo  {journal} {Phys. Rev. Lett.}\ }\textbf {\bibinfo {volume} {114}},\ \bibinfo {pages} {220405} (\bibinfo {year} {2015})}\BibitemShut {NoStop}%
\bibitem [{\citenamefont {Paris}(2015)}]{Paris2016-thermometry}%
  \BibitemOpen
  \bibfield  {author} {\bibinfo {author} {\bibfnamefont {M.~G.~A.}\ \bibnamefont {Paris}},\ }\bibfield  {title} {\bibinfo {title} {Achieving the landau bound to precision of quantum thermometry in systems with vanishing gap},\ }\href {https://doi.org/10.1088/1751-8113/49/3/03LT02} {\bibfield  {journal} {\bibinfo  {journal} {J. Phys. A: Math. Theor.}\ }\textbf {\bibinfo {volume} {49}},\ \bibinfo {pages} {03LT02} (\bibinfo {year} {2015})}\BibitemShut {NoStop}%
\bibitem [{\citenamefont {Hofer}\ \emph {et~al.}(2017)\citenamefont {Hofer}, \citenamefont {Brask}, \citenamefont {Perarnau-Llobet},\ and\ \citenamefont {Brunner}}]{Hofer2017-thermometry}%
  \BibitemOpen
  \bibfield  {author} {\bibinfo {author} {\bibfnamefont {P.~P.}\ \bibnamefont {Hofer}}, \bibinfo {author} {\bibfnamefont {J.~B.}\ \bibnamefont {Brask}}, \bibinfo {author} {\bibfnamefont {M.}~\bibnamefont {Perarnau-Llobet}},\ and\ \bibinfo {author} {\bibfnamefont {N.}~\bibnamefont {Brunner}},\ }\bibfield  {title} {\bibinfo {title} {Quantum thermal machine as a thermometer},\ }\href {https://doi.org/10.1103/PhysRevLett.119.090603} {\bibfield  {journal} {\bibinfo  {journal} {Phys. Rev. Lett.}\ }\textbf {\bibinfo {volume} {119}},\ \bibinfo {pages} {090603} (\bibinfo {year} {2017})}\BibitemShut {NoStop}%
\bibitem [{\citenamefont {De~Pasquale}\ and\ \citenamefont {Stace}(2018)}]{DePasquale2018-thermometry}%
  \BibitemOpen
  \bibfield  {author} {\bibinfo {author} {\bibfnamefont {A.}~\bibnamefont {De~Pasquale}}\ and\ \bibinfo {author} {\bibfnamefont {T.~M.}\ \bibnamefont {Stace}},\ }\bibinfo {title} {Quantum thermometry},\ in\ \href {https://doi.org/10.1007/978-3-319-99046-0_21} {\emph {\bibinfo {booktitle} {Thermodynamics in the Quantum Regime: Fundamental Aspects and New Directions}}},\ \bibinfo {editor} {edited by\ \bibinfo {editor} {\bibfnamefont {F.}~\bibnamefont {Binder}}, \bibinfo {editor} {\bibfnamefont {L.~A.}\ \bibnamefont {Correa}}, \bibinfo {editor} {\bibfnamefont {C.}~\bibnamefont {Gogolin}}, \bibinfo {editor} {\bibfnamefont {J.}~\bibnamefont {Anders}},\ and\ \bibinfo {editor} {\bibfnamefont {G.}~\bibnamefont {Adesso}}}\ (\bibinfo  {publisher} {Springer International Publishing},\ \bibinfo {address} {Cham},\ \bibinfo {year} {2018})\ pp.\ \bibinfo {pages} {503--527}\BibitemShut {NoStop}%
\bibitem [{\citenamefont {Mehboudi}\ \emph {et~al.}(2019)\citenamefont {Mehboudi}, \citenamefont {Sanpera},\ and\ \citenamefont {Correa}}]{Mehboudi2019-thermometry}%
  \BibitemOpen
  \bibfield  {author} {\bibinfo {author} {\bibfnamefont {M.}~\bibnamefont {Mehboudi}}, \bibinfo {author} {\bibfnamefont {A.}~\bibnamefont {Sanpera}},\ and\ \bibinfo {author} {\bibfnamefont {L.~A.}\ \bibnamefont {Correa}},\ }\bibfield  {title} {\bibinfo {title} {Thermometry in the quantum regime: recent theoretical progress},\ }\href {https://doi.org/10.1088/1751-8121/ab2828} {\bibfield  {journal} {\bibinfo  {journal} {J. Phys. A: Math. Theor.}\ }\textbf {\bibinfo {volume} {52}},\ \bibinfo {pages} {303001} (\bibinfo {year} {2019})}\BibitemShut {NoStop}%
\bibitem [{\citenamefont {Potts}\ \emph {et~al.}(2019)\citenamefont {Potts}, \citenamefont {Brask},\ and\ \citenamefont {Brunner}}]{Potts2019-thermometry}%
  \BibitemOpen
  \bibfield  {author} {\bibinfo {author} {\bibfnamefont {P.~P.}\ \bibnamefont {Potts}}, \bibinfo {author} {\bibfnamefont {J.~B.}\ \bibnamefont {Brask}},\ and\ \bibinfo {author} {\bibfnamefont {N.}~\bibnamefont {Brunner}},\ }\bibfield  {title} {\bibinfo {title} {Fundamental limits on low-temperature quantum thermometry with finite resolution},\ }\href {https://doi.org/10.22331/q-2019-07-09-161} {\bibfield  {journal} {\bibinfo  {journal} {{Quantum}}\ }\textbf {\bibinfo {volume} {3}},\ \bibinfo {pages} {161} (\bibinfo {year} {2019})}\BibitemShut {NoStop}%
\bibitem [{\citenamefont {J\o{}rgensen}\ \emph {et~al.}(2020)\citenamefont {J\o{}rgensen}, \citenamefont {Potts}, \citenamefont {Paris},\ and\ \citenamefont {Brask}}]{Jorgensen2020-thermometry}%
  \BibitemOpen
  \bibfield  {author} {\bibinfo {author} {\bibfnamefont {M.~R.}\ \bibnamefont {J\o{}rgensen}}, \bibinfo {author} {\bibfnamefont {P.~P.}\ \bibnamefont {Potts}}, \bibinfo {author} {\bibfnamefont {M.~G.~A.}\ \bibnamefont {Paris}},\ and\ \bibinfo {author} {\bibfnamefont {J.~B.}\ \bibnamefont {Brask}},\ }\bibfield  {title} {\bibinfo {title} {Tight bound on finite-resolution quantum thermometry at low temperatures},\ }\href {https://doi.org/10.1103/PhysRevResearch.2.033394} {\bibfield  {journal} {\bibinfo  {journal} {Phys. Rev. Res.}\ }\textbf {\bibinfo {volume} {2}},\ \bibinfo {pages} {033394} (\bibinfo {year} {2020})}\BibitemShut {NoStop}%
\bibitem [{\citenamefont {Hovhannisyan}\ \emph {et~al.}(2021)\citenamefont {Hovhannisyan}, \citenamefont {J\o{}rgensen}, \citenamefont {Landi}, \citenamefont {Alhambra}, \citenamefont {Brask},\ and\ \citenamefont {Perarnau-Llobet}}]{Hovhannisyan2021-thermometry}%
  \BibitemOpen
  \bibfield  {author} {\bibinfo {author} {\bibfnamefont {K.~V.}\ \bibnamefont {Hovhannisyan}}, \bibinfo {author} {\bibfnamefont {M.~R.}\ \bibnamefont {J\o{}rgensen}}, \bibinfo {author} {\bibfnamefont {G.~T.}\ \bibnamefont {Landi}}, \bibinfo {author} {\bibfnamefont {A.~M.}\ \bibnamefont {Alhambra}}, \bibinfo {author} {\bibfnamefont {J.~B.}\ \bibnamefont {Brask}},\ and\ \bibinfo {author} {\bibfnamefont {M.}~\bibnamefont {Perarnau-Llobet}},\ }\bibfield  {title} {\bibinfo {title} {Optimal quantum thermometry with coarse-grained measurements},\ }\href {https://doi.org/10.1103/PRXQuantum.2.020322} {\bibfield  {journal} {\bibinfo  {journal} {PRX Quantum}\ }\textbf {\bibinfo {volume} {2}},\ \bibinfo {pages} {020322} (\bibinfo {year} {2021})}\BibitemShut {NoStop}%
\bibitem [{\citenamefont {Rubio}\ \emph {et~al.}(2021)\citenamefont {Rubio}, \citenamefont {Anders},\ and\ \citenamefont {Correa}}]{Rubio2021-thermometry}%
  \BibitemOpen
  \bibfield  {author} {\bibinfo {author} {\bibfnamefont {J.}~\bibnamefont {Rubio}}, \bibinfo {author} {\bibfnamefont {J.}~\bibnamefont {Anders}},\ and\ \bibinfo {author} {\bibfnamefont {L.~A.}\ \bibnamefont {Correa}},\ }\bibfield  {title} {\bibinfo {title} {Global quantum thermometry},\ }\href {https://doi.org/10.1103/PhysRevLett.127.190402} {\bibfield  {journal} {\bibinfo  {journal} {Phys. Rev. Lett.}\ }\textbf {\bibinfo {volume} {127}},\ \bibinfo {pages} {190402} (\bibinfo {year} {2021})}\BibitemShut {NoStop}%
\bibitem [{\citenamefont {Jahnke}\ \emph {et~al.}(2011)\citenamefont {Jahnke}, \citenamefont {Lan\'ery},\ and\ \citenamefont {Mahler}}]{PhysRevE.83.011109}%
  \BibitemOpen
  \bibfield  {author} {\bibinfo {author} {\bibfnamefont {T.}~\bibnamefont {Jahnke}}, \bibinfo {author} {\bibfnamefont {S.}~\bibnamefont {Lan\'ery}},\ and\ \bibinfo {author} {\bibfnamefont {G.}~\bibnamefont {Mahler}},\ }\bibfield  {title} {\bibinfo {title} {Operational approach to fluctuations of thermodynamic variables in finite quantum systems},\ }\href {https://doi.org/10.1103/PhysRevE.83.011109} {\bibfield  {journal} {\bibinfo  {journal} {Phys. Rev. E}\ }\textbf {\bibinfo {volume} {83}},\ \bibinfo {pages} {011109} (\bibinfo {year} {2011})}\BibitemShut {NoStop}%
\bibitem [{\citenamefont {Born}(1926)}]{Born1926}%
  \BibitemOpen
  \bibfield  {author} {\bibinfo {author} {\bibfnamefont {M.}~\bibnamefont {Born}},\ }\bibfield  {title} {\bibinfo {title} {Zur quantenmechanik der sto{\ss}vorg{\"a}nge},\ }\href {https://doi.org/10.1007/BF01397477} {\bibfield  {journal} {\bibinfo  {journal} {Zeitschrift f{\"u}r Physik}\ }\textbf {\bibinfo {volume} {37}},\ \bibinfo {pages} {863} (\bibinfo {year} {1926})}\BibitemShut {NoStop}%
\bibitem [{\citenamefont {Rao}(1945)}]{MR15748}%
  \BibitemOpen
  \bibfield  {author} {\bibinfo {author} {\bibfnamefont {C.~R.}\ \bibnamefont {Rao}},\ }\bibfield  {title} {\bibinfo {title} {Information and the accuracy attainable in the estimation of statistical parameters},\ }\href@noop {} {\bibfield  {journal} {\bibinfo  {journal} {Bull. Calcutta Math. Soc.}\ }\textbf {\bibinfo {volume} {37}},\ \bibinfo {pages} {81} (\bibinfo {year} {1945})}\BibitemShut {NoStop}%
\bibitem [{\citenamefont {Cram\'er}(1946)}]{a61aa5fe-d74a-3133-bed2-f35c3c555015}%
  \BibitemOpen
  \bibfield  {author} {\bibinfo {author} {\bibfnamefont {H.}~\bibnamefont {Cram\'er}},\ }\href {http://www.jstor.org/stable/j.ctt1bpm9r4} {\emph {\bibinfo {title} {Mathematical Methods of Statistics}}}\ (\bibinfo  {publisher} {Princeton University Press},\ \bibinfo {year} {1946})\BibitemShut {NoStop}%
\bibitem [{\citenamefont {Kay}(1993)}]{10.5555/151045}%
  \BibitemOpen
  \bibfield  {author} {\bibinfo {author} {\bibfnamefont {S.~M.}\ \bibnamefont {Kay}},\ }\href@noop {} {\emph {\bibinfo {title} {Fundamentals of statistical signal processing: estimation theory}}}\ (\bibinfo  {publisher} {Prentice-Hall, Inc.},\ \bibinfo {address} {USA},\ \bibinfo {year} {1993})\BibitemShut {NoStop}%
\bibitem [{\citenamefont {Wootters}(1981)}]{PhysRevD.23.357}%
  \BibitemOpen
  \bibfield  {author} {\bibinfo {author} {\bibfnamefont {W.~K.}\ \bibnamefont {Wootters}},\ }\bibfield  {title} {\bibinfo {title} {Statistical distance and hilbert space},\ }\href {https://doi.org/10.1103/PhysRevD.23.357} {\bibfield  {journal} {\bibinfo  {journal} {Phys. Rev. D}\ }\textbf {\bibinfo {volume} {23}},\ \bibinfo {pages} {357} (\bibinfo {year} {1981})}\BibitemShut {NoStop}%
\bibitem [{\citenamefont {Breuer}\ and\ \citenamefont {Petruccione}(2007)}]{10.1093/acprof:oso/9780199213900.001.0001}%
  \BibitemOpen
  \bibfield  {author} {\bibinfo {author} {\bibfnamefont {H.-P.}\ \bibnamefont {Breuer}}\ and\ \bibinfo {author} {\bibfnamefont {F.}~\bibnamefont {Petruccione}},\ }\href {https://doi.org/10.1093/acprof:oso/9780199213900.001.0001} {\emph {\bibinfo {title} {The Theory of Open Quantum Systems}}}\ (\bibinfo  {publisher} {Oxford University Press},\ \bibinfo {year} {2007})\BibitemShut {NoStop}%
\bibitem [{\citenamefont {Diamond}\ and\ \citenamefont {Boyd}(2016)}]{diamond2016cvxpy}%
  \BibitemOpen
  \bibfield  {author} {\bibinfo {author} {\bibfnamefont {S.}~\bibnamefont {Diamond}}\ and\ \bibinfo {author} {\bibfnamefont {S.}~\bibnamefont {Boyd}},\ }\bibfield  {title} {\bibinfo {title} {{CVXPY}: {A} {P}ython-embedded modeling language for convex optimization},\ }\href {http://jmlr.org/papers/v17/15-408.html} {\bibfield  {journal} {\bibinfo  {journal} {J. Mach. Learn. Res.}\ }\textbf {\bibinfo {volume} {17}},\ \bibinfo {pages} {1} (\bibinfo {year} {2016})}\BibitemShut {NoStop}%
\bibitem [{\citenamefont {Agrawal}\ \emph {et~al.}(2018)\citenamefont {Agrawal}, \citenamefont {Verschueren}, \citenamefont {Diamond},\ and\ \citenamefont {Boyd}}]{agrawal2018rewriting}%
  \BibitemOpen
  \bibfield  {author} {\bibinfo {author} {\bibfnamefont {A.}~\bibnamefont {Agrawal}}, \bibinfo {author} {\bibfnamefont {R.}~\bibnamefont {Verschueren}}, \bibinfo {author} {\bibfnamefont {S.}~\bibnamefont {Diamond}},\ and\ \bibinfo {author} {\bibfnamefont {S.}~\bibnamefont {Boyd}},\ }\bibfield  {title} {\bibinfo {title} {A rewriting system for convex optimization problems},\ }\href {https://doi.org/10.1080/23307706.2017.1397554} {\bibfield  {journal} {\bibinfo  {journal} {JCD}\ }\textbf {\bibinfo {volume} {5}},\ \bibinfo {pages} {42} (\bibinfo {year} {2018})}\BibitemShut {NoStop}%
\bibitem [{\citenamefont {{MOSEK ApS}}(2024)}]{mosek}%
  \BibitemOpen
  \bibfield  {author} {\bibinfo {author} {\bibnamefont {{MOSEK ApS}}},\ }\href {https://www.mosek.com} {\emph {\bibinfo {title} {The MOSEK Optimization Toolbox, Version 11.0.25}}} (\bibinfo {year} {2024})\BibitemShut {NoStop}%
\end{thebibliography}%
\appendix
\section{Lindblad operators for strong coupling regime} \label{app}
The frequencies $\{\omega_j\}$ are given as
\begin{align}
    \omega_1=E_1,\omega_2=E_1+g,\omega_3=E_1-g,\nonumber\\
     \omega_4=E_2,\omega_5=E_2+g,\omega_6=E_2-g,\nonumber\\
      \omega_7=E_3,\omega_8=E_3+g,\omega_9=E_3-g.
\end{align}
The Lindblad operators are 
\begin{equation}
\begin{aligned}
\mathcal{L}_1(E_1) &= \ket{011}\bra{111} + \ket{000}\bra{100}, \\
\mathcal{L}_2(E_1 + g) &= \frac{1}{\sqrt{2}} \left( \ket{100}\bra{+} - \ket{-}\bra{110} \right), \\
\mathcal{L}_3(E_1 - g) &= \frac{1}{\sqrt{2}} \left( \ket{+}\bra{110} + \ket{100}\bra{-} \right), \\[4pt]
\mathcal{L}_4(E_2) &= \ket{100}\bra{110} + \ket{001}\bra{011}, \\
\mathcal{L}_5(E_2 + g) &= \frac{1}{\sqrt{2}} \left( \ket{000}\bra{+} - \ket{-}\bra{111} \right), \\
\mathcal{L}_6(E_2 - g) &= \frac{1}{\sqrt{2}} \left( \ket{+}\bra{111} - \ket{000}\bra{-} \right), \\[4pt]
\mathcal{L}_7(E_3) &= \ket{110}\bra{111} + \ket{000}\bra{001}, \\
\mathcal{L}_8(E_3 + g) &= \frac{1}{\sqrt{2}} \left( \ket{100}\bra{+} - \ket{-}\bra{011} \right), \\
\mathcal{L}_9(E_3 - g) &= \frac{1}{\sqrt{2}} \left( \ket{+}\bra{011} + \ket{100}\bra{-}. \right)
\end{aligned}
\end{equation}
Here we have $\ket{\pm}=\frac{1}{\sqrt{2}}\left(\ket{101}\pm\ket{010}\right)$. The remaining nine operators for the inverse processes are obtained from the relation $\mathcal{L}_j(-\omega_j)=\mathcal{L}^{\dagger}_j(\omega_j)$.
\end{document}